\documentclass[10pt]{article}
\usepackage{fullpage,citesort,epsfig,psfrag,graphics,amsmath,amsbsy,amssymb}
\usepackage{caption}

\newcommand{\bea}{\begin{eqnarray}}
\newcommand{\eea}{\end{eqnarray}}
\newcommand{\beq}{\begin{equation}}
\newcommand{\eeq}{\end{equation}}

\newcommand{\bfs}{\boldsymbol}

\newcommand{\be}{\begin{equation}}
\newcommand{\ee}{\end{equation}}
\newcommand{\bq}{\begin{eqnarray}}
\newcommand{\eq}{\end{eqnarray}}

\def\math{\mathsurround=0pt }
\def\leftrightarrowfill{$\math \mathord\leftarrow \mkern-6mu 
 \cleaders\hbox{$\mkern-2mu \mathord- \mkern-2mu$}\hfill
 \mkern-6mu \mathord\rightarrow$}
\def\overleftrightarrow#1{\vbox{\ialign{##\crcr
     \leftrightarrowfill\crcr\noalign{\kern-1pt\nointerlineskip}
     $\hfil\displaystyle{#1}\hfil$\crcr}}}

\def\m@th{\mathsurround=0pt }
\def\leftrightarrowfill{$\m@th \mathord\leftarrow \mkern-6mu
 \cleaders\hbox{$\mkern-2mu \mathord- \mkern-2mu$}\hfill
 \mkern-6mu \mathord\rightarrow$}
\def\overleftrightarrow#1{\vbox{\ialign{##\crcr
     \leftrightarrowfill\crcr\noalign{\kern-1pt\nointerlineskip}
     $\hfil\displaystyle{#1}\hfil$\crcr}}}
\begin{document}
\setlength{\captionmargin}{36pt}
\begin{titlepage}
\begin{flushright}
UFIFT-HEP-08-6\\
\end{flushright}

\vskip 3cm

\begin{center}
\begin{Large}
{\bf Classical Worldsheets for String Scattering\\ 
on Flat and AdS Spacetime
}
\end{Large}
\end{center}
\vskip 2cm
\begin{center}
{\large 
Charles M. Sommerfield\footnote{E-mail  address: 
{\tt sommerfield@phys.ufl.edu}} and 
Charles B. Thorn\footnote{E-mail  address: {\tt thorn@phys.ufl.edu}}
}
\vskip0.20cm
{\it Institute for Fundamental Theory\\
Department of Physics, University of Florida,
Gainesville FL 32611}\\

\vskip 1.0cm
\end{center}

\begin{abstract}
\noindent We present a study of the worldsheets that 
describe the classical limit of various string scattering processes. 
Our main focus is on string scattering in AdS spacetime
because of its relation via the AdS/CFT correspondence to 
gluon scattering in ${\cal N}=4$ supersymmetric Yang-Mills
theory. But we also consider analogous processes in flat Minkowski
spacetime which we compare to the AdS case. In addition to
scattering of string by string we also find and study worldsheets
describing the scattering of a string by external sources.
\end{abstract}
\vfill
\end{titlepage}
\section{Introduction}
The exact tree amplitudes for the scattering of strings in flat spacetime 
can be obtained in many ways. For example, the path integral for the sum over
worldsheets can be completely evaluated by choosing a parametrization
that linearizes the worldsheet equations. Because of this, one doesn't
really need a visualization of the ``classical'' worldsheet that would
represent a saddle-point approximation to the path integral. But for
string scattering on non-trivial backgrounds, for which an exact 
evaluation is not possible,
the saddle-point approximation is frequently the only available approach
to calculations.
Then it becomes important to understand the physical meaning of 
various classical worldsheet solutions.

Because of the AdS/CFT correspondence \cite{maldacena}, 
the scattering of field quanta of a conformal quantum field theory is
dual to string scattering on an Anti-deSitter spacetime, and the
strong 't Hooft coupling limit on the field side is dual to the classical limit
on the string side.
For example, Alday and Maldacena (AM) \cite{aldaym4pt,aldaymnpt}
have employed this duality 
to calculate the scattering of gluons and their superpartners in ${\cal N}=4$
supersymmetric Yang-Mills theory in the limit of strong coupling.
This limit is described by the classical dynamics of string
moving in an Anti-deSitter (AdS) spacetime. To describe the
scattering one needs to find the classical worldsheet in Poincar\'e
coordinates $x^\mu(\xi), z(\xi)$, with its boundary
at $z=\infty$.\footnote{Here $\xi=(\xi^0,\xi^1)$ are the 2D worldsheet coordinates of the string.}  Actually, it is easier to work with
T-dual coordinates \cite{thooftlargen,bardakcit}
$y^\mu(\xi)$ and to use $r=R^2/z$
instead of $z$. These coordinates turn out to be the Poincar\'e coordinates
of another AdS manifold. In terms of them, the boundary of the
desired worldsheet is at $r=0$, and, for the scattering of $n$
ground state strings, the boundary conditions are
simply that the worldsheet end at $r=0$ on a closed polygon
of $n$ lightlike line segments, given by the momenta
of the $n$ external particles \cite{aldaym4pt}.

One can also use T-dual coordinates to describe string scattering in flat
spacetime. This is usually not done, but we think it a useful 
and interesting exercise to examine the T-dual description of worldsheets 
in this case as well. In flat spacetime,
choosing conformally flat worldsheet coordinates 
linearizes the equations of motion,
and the worldsheet solution can be expressed in 
terms of  Green functions, with appropriate boundary conditions,
for the 2D Laplacian.  Studying the solutions in
flat spacetime helps clarify the physical nature of the boundary conditions
used by AM in the worldsheet solutions on AdS applied to gluon
scattering at strong coupling.
For example, the scattering amplitudes can be explicitly
related to finite time transition amplitudes in the lightcone
gauge, and therefore can be obtained from a reduction formalism. 
In particular, it
is interesting to compare the flat and AdS worldsheets as a way
to illuminate the very different physics in the two cases: in the
flat case the string tension $T_0$ is a constant implying that
the energy of a stretched string is proportional to its length,
whereas in the AdS case the effective tension depends on
$z$, $T_{\rm eff}(z)$, dropping to 0 at $z=\infty$ (or $r=0$).
The vanishing tension in the second case implies a continuous
mass spectrum of string excitations which complicates the application
of the reduction formalism (as with all theories with massless
particles).

Thus in this article
we set out to study the worldsheets appropriate to various scattering
processes in both flat and AdS spacetime. 
In AdS, exact solutions for string motion are hard to come by. We look at the
solutions used by AM for scattering of gluons
by gluons, which were constructed from earlier ones
found by Kruczenski \cite{kruczenski}, as well as a more general
solution involving a gluon scattering off some kind of
external source. This latter process, however, does not look like a gluon
scattering from a heavy quark represented on AdS by a string 
stretching from a point at $z=0$ out to $z=\infty$, a process 
we had hoped to understand.  Instead, for this solution the 
end of the string interacting with the source is at $z=\infty$ ($r=0$).

In all cases we display the solutions in the T-dual coordinates
$y^\mu(\sigma,\tau)$ and, in the case of AdS, $r$. 
These coordinates are particularly nice
because the interesting worldsheets are of finite extent in them,
and they can be displayed graphically in a striking way.
In the case of on-shell gluon scattering, the Minkowski
part of the worldsheet in AdS is at first glance qualitatively very similar
to the flat space worldsheet.
The very different physics can only be understood after 
considering the quantitative differences due to the dynamics
of the AdS radius $r$. 
We also find worldsheets for a gluon scattering from a 0-brane
or a 1-brane in flat spacetime. These can be compared to the more general
solution in AdS mentioned at the end of the previous paragraph,
and we study them to gain insight into the physics of the latter
solution.

The rest of the paper is organized as follows. The next two
sections provide background for the subsequent studies. In section 2
we discuss the T-duality transformation from the point of view of the
phase-space action principle. Section 3 gives a brief discussion of
the properties of the ``single gluon'' defined by AM using a
3-brane at fixed finite $z_{IR}$. 
We begin those studies in section 4 which discusses the 
worldsheets appropriate to $n$ string scattering in flat space.
All our solutions in AdS$_5$ are constructed from exact solutions 
on an AdS$_3$ submanifold that are found and described in section 5.
Then section 6 discusses in this general framework the AdS$_5$
solution for 4 gluon scattering used by AM. Section 7 discusses a
new solution in AdS$_5$ that describes a single gluon scattering from
an external source. Section 8 employs
lightcone gauge to compare the
physics of the AdS 4-string solutions to that of the flat space solutions. 
Finally, to gain further insight into
such external source scattering processes we study the worldsheets
for scattering of a string off 0-brane and 1-brane sources in flat space
in section 9. 
We include one appendix on useful facts about various covariant and 
lightcone worldsheet parametrizations, and a second on
the relation between conformal transformations in 4 dimensional
spacetime and isometry transformations of AdS$_5$.

\section{T-duality and the Phase Space Worldsheet Action}
We choose coordinates for AdS spacetime so that the line element 
is $ds^2={R^2}(dx_\mu dx^\mu+dz^2)/z^2$. Then the worldsheet action
for a string moving on AdS$_5$ is in Nambu-Goto form\footnote{in this section we use  $^.$ and $^\prime$  to represent ${\partial\over\partial\xi^0}$  and ${\partial\over\partial\xi^1}$, respectively.}
\bea
S_{NG}&=& -T_0\int d^2\xi{R^2\over z^2}\sqrt{({\dot x}\cdot x^\prime+{\dot z}
z^\prime)^2-({\dot x}^2+{\dot z}^2)(x^{\prime 2}+z^{\prime 2})}.
\label{xnambugoto}
\eea
With the introduction of a dynamical worldsheet metric $g_{\alpha\beta}$ we get an equivalent form
\bea S_{ws}\equiv\int
d^2\xi{\cal L}=-{T_0\over2}\int d^2\xi\sqrt{g}g^{\alpha\beta}
{R^2\over z^2}(\partial_\alpha x\cdot\partial_\beta x+\partial_\alpha z
\partial_\beta z).
\eea
For ${\cal N}=4$ super Yang-Mills, the AdS/CFT correspondence gives
$T_0R^2=\sqrt{g^2N_c/4\pi^2}=\sqrt{\lambda}$.
The momenta conjugate to $(x^\mu, z)$ are $({\cal P}^\mu,\Pi)
=-R^2T_0\sqrt{g}g^{0\beta}
\partial_\beta (x^\mu,z)/z^2$. Then the canonical hamiltonian density is
\bea
{\cal H}&=&{\dot x}\cdot{\cal P}+{\dot z}\Pi-{\cal L}
\nonumber\\
&=&{\kappa\over2}\left({z^2\over T_0R^2}({\cal P}^2+\Pi^2)
+{T_0R^2\over z^2}(x^{\prime2}+z^{\prime2})\right)+\mu(x^\prime\cdot{\cal P}
+z^\prime\Pi),\\
\kappa&=&-{1\over g^{00}\sqrt{g}},\qquad \mu=-{g^{01}\over g^{00}}.
\eea
If, instead,  the canonical momentum is obtained from the Nambu-Goto action
(\ref{xnambugoto}), ${\cal H}$ and 
the coefficients of $\kappa$, $\mu$ are identically
zero.This represents phase space constraints for which $\kappa$, $\mu$
are then introduced as  Lagrange multipliers.
In either case, the phase space Lagrange density is 
\cite{goddardgrt,metsaevtt}
\bea
{\cal L}
&=&{\dot x}\cdot{\cal P}+{\dot z}\Pi
-{\kappa\over2}\left({z^2\over T_0R^2}({\cal P}^2+\Pi^2)
+{T_0R^2\over z^2}(x^{\prime2}+z^{\prime2})\right)
-\mu(x^\prime\cdot{\cal P}+z^\prime\Pi)\\
&\to&{\dot x}\cdot{\cal P}
-{\kappa\over2}\left({z^2\over T_0R^2}{\cal P}^2
+{T_0R^2\over z^2}x^{\prime2}\right)-\mu x^\prime\cdot{\cal P}
+{R^2T_0\over2z^2}\left[
{({\dot z}-\mu z^\prime)^2\over \kappa}-\kappa z^{\prime2}\right],
\eea
where in the second line we have eliminated $\Pi$ using its purely
algebraic equation of motion.
To discuss duality we simply replace ${\cal P}=q^\prime\equiv T_0y^\prime$:
\bea
{\cal L}
&=&T_0\left({\dot x}\cdot y^\prime
-{\kappa\over2}\left({z^2\over R^2}y^{\prime2}
+{R^2\over z^2}x^{\prime2}\right)-\mu(x^\prime\cdot
y^\prime)+{R^2\over2z^2}\left[
{({\dot z}-\mu z^\prime)^2\over \kappa}-\kappa z^{\prime2}\right]\right)
\label{manduality}
\eea
and we immediately see a symmetry up to surface terms 
under $x\leftrightarrow q/T_0=y$, $z\to R^2/z$. This is
T-duality. In the form written
in (\ref{manduality}), ${\cal L}$ 
gives a good action principle for the variable $x$. Note that
the absence of ${\dot y}$ in ${\cal L}$ implies a gauge symmetry
under $y\to y + f(\xi^0)$. Varying $y$ with open string topology
implies $y^\prime = R^2(\dot{x}-\mu x^\prime)/(\kappa z^2)$.
Then varying $x$ gives
\bea
{\dot y}^\prime=\left[{\kappa R^2\over z^2}x^\prime
+\mu y^\prime\right]^\prime,\qquad \left[{\kappa R^2\over z^2}x^\prime
+\mu y^\prime\right]_{\rm ends}=0.
\eea
Thus this version of the action principle
implies the duality transformation in the form
\bea
y^\prime={R^2\over\kappa z^2}(\dot{x}-\mu x^\prime),\qquad
{\dot y}={\kappa R^2\over z^2}x^\prime
+\mu y^\prime+f(\xi^0)={\kappa R^2\over z^2}x^\prime
+\mu {R^2\over\kappa z^2}(\dot{x}-\mu x^\prime)+f(\xi^0)
\eea
with undetermined $f(\xi^0)$, which can, if desired, 
be removed by a gauge transformation, giving
\bea
y^\prime={R^2\over\kappa z^2}(\dot{x}-\mu x^\prime),\qquad
{\dot y}-\mu y^\prime={\kappa R^2\over z^2}x^\prime.
\label{genduality}
\eea
 Also notice that the
free-end boundary conditions on $x$ imply ${\dot y}_{\rm ends}=f(\xi^0)$
which implies that $d{\Delta{y}}/d\xi^0=0$, where $\Delta y$ is the difference
in the values of $y$ at the two ends. In a gauge where $f(\xi^0)=0$
$y$ satisfies Dirichlet conditions ${\dot y}_{\rm ends}=0$ at the ends.

Next consider Dirichlet conditions on $x$, ${\dot x}_{\rm ends}=0$.
Then
\bea
{y^\prime}_{\rm ends}=-{R^2\mu\over\kappa z^2}x^\prime
=-{\mu\over \kappa^2}({\dot y}-\mu y^\prime-f(\xi^0))
\eea
which, in a gauge where $f(\xi^0)=0$, are precisely analogous to the
free end condition on $x$. This condition would follow as
a free end condition on $y$ from the action obtained from
(\ref{manduality}) by substituting ${\dot x}\cdot y^\prime
\to{\dot y}\cdot x^\prime$.
Indeed, this substitution must be made to get a good action 
principle for the variable $y$. We find
\bea
{\dot x}\cdot y^\prime-{\dot y}\cdot x^\prime={\partial\over\partial\xi^0}
({x}\cdot y^\prime)-{\partial\over\partial\sigma}
({x}\cdot {\dot y})=-{\partial\over\partial\xi^0}
({y}\cdot x^\prime)+{\partial\over\partial\sigma}
({y}\cdot {\dot x})
\eea
so the substitution is achieved by adding a surface term to the action.
Doing this and reversing the logic, we conclude that the Nambu-Goto
form of the action for the T-dual variables $y$ is
\bea
{\tilde S}_{NG}&=& -T_0\int d^2\xi{R^2\over r^2}
\sqrt{({\dot y}\cdot y^\prime+{\dot r}
r^\prime)^2-({\dot y}^2+{\dot r}^2)(y^{\prime 2}+r^{\prime 2})}.
\label{ynambugoto}
\eea
Either action principle has identical consequences for the
dual pair $x, y$ modulo the gauge invariance appropriate to the
action chosen.

For closed-string topology, these duality results will 
follow if we only require
$y^\prime$ to be periodic, allowing $y$ to shift by a function of
$\xi^0$ upon traversing a period: that is $y$ is required
to be periodic up to a gauge transformation.

Finally we determine the worldsheet metric in terms of the
$x, z$ or the $y= q/T_0, r\equiv R^2/z$ by varying 
the action with respect to $\kappa,\mu$, and using the duality
relations (with $f(\xi^0)=0$):
\bea
\mu&=& {{\dot x}\cdot x^\prime+{\dot z}\cdot z^\prime
\over x^{\prime2}+ z^{\prime2}}\ =\ 
{{\dot y}\cdot y^\prime+{\dot r}\cdot r^\prime
\over y^{\prime2}+ r^{\prime2}},\\
\kappa^2 &=& {({\dot x}\cdot x^\prime+{\dot z} z^\prime)^2
-({\dot x}^2+{\dot z}^{2})(x^{\prime2}+ z^{\prime2})\over
(x^{\prime2}+ z^{\prime2})^2}\ =\ 
{({\dot y}\cdot y^\prime+{\dot r} r^\prime)^2
-({\dot y}^2+{\dot r}^{2})(y^{\prime2}+ r^{\prime2})\over
(y^{\prime2}+ r^{\prime2})^2}.
\eea
\section{Defining the strongly coupled ``gluon''}
Alday and Maldacena were principally interested in using the AdS/CFT
correspondence to study the scattering of gluons in ${\cal N}=4$
supersymmetric Yang-Mills theory. 
For purposes of comparing to work of Bern, Dixon, and Smirnov (BDS)
\cite{bernds}
they needed to define on-shell amplitudes with a definite number of gluons
at strong coupling. 
Strictly speaking all these amplitudes are zero due to infrared divergences: 
the non-vanishing physical observables are jets with an indefinite number
of gluons. Thus an IR cutoff that enables the definition
of single gluon states is needed. 
One practical infrared cutoff is to use dimensional
regularization, continuing to $D=4+\epsilon$.
This is used by BDS in their perturbative calculations,
which establish the definition of single gluons using
the dressed gluon propagator. 

However the notion of a dressed single-gluon propagator is
obscure in the string description. Instead AM define single
gluons as open AdS strings with ends moving on a D3 brane at fixed
$z=z_{IR}<\infty$. They then take $z_{IR}\to\infty$ after
continuing to $D=4+\epsilon$ dimensions. 
In the string description the IR divergences are
due to the huge degeneracy of single open-string mass eigenstates
when the effective rest tension vanishes as $z_{IR}\to\infty$.\footnote{
At finite $z_{IR}$ a string with ends fixed at finite $z_{IR}$ has
an energy gap at $N=\infty$ 
(see \cite{maldacenaqqbar,callang,klebanovmt}).}
This is
the string dual to the huge degeneracy in the field description 
due to the presence of many massless gluons. Fixing the ends of the
open string representing a gluon to $z=z_{IR}$ breaks this
degeneracy by introducing an energy cost to stretching an open
string with ends fixed to the brane. It is not a confining
energy cost (it stays finite at infinite separation of the 
string ends), but a finite classical energy cost is enough to discretize
the semi-classical energy near zero, producing a gap which
enables the lowest mass eigenstate, which is the ${\cal N}=4$ supermultiplet 
containing the gluon, to be filtered out by the reduction
formalism. 

At strong coupling the string dynamics is
classical, and it is significant that the unstretched classical open
string (i.e. $x^\prime=0$ for all $\xi^1$), corresponding to the ground
state of the string and identified with the gluon, 
solves the full set of classical AdS string equations at 
constant $z$: ${\dot z}=z^\prime=0$. Such 
a solution requires ${\dot x}\cdot{\dot x}=0$,
i.e. the unstretched open string moves freely on the brane
at the speed of light---it is a massless point particle. In T-dual variables
this condition reads $y^\prime\cdot y^\prime=0$. In other
words the reduction formalism applied at finite $z_{IR}$ (which
supplies a mass gap)
will require the classical worldsheet
in T-dual space for gluon scattering amplitudes
to end at $z=z_{IR}$ on light-like line segments corresponding
to the momenta of the on-shell gluons. This is precisely the
prescription employed by Alday and Maldacena. 

If $z_{IR}$ is finite the open string modeling the single gluon
has internal structure that can be analyzed semi-classically.
Indeed, the model is mathematically equivalent to the models
of $Q{\bar Q}$ mesons studied in the context of AdS/CFT inspired
models of QCD. For example, in \cite{kruczenskimmw} the rotational motion
of such an open string, with ends fixed to a 3-brane
at fixed $z$ is solved numerically. An interesting outcome of that
work is the leading Regge trajectory $J(E^2)$, described by
a straight open string rotating rigidly about its midpoint. 
The asymptotic behavior of the trajectory at small $E$ and near
ionization threshold, which we translate to the gluon parameters,
was obtained analytically.
For $E\sim0$, it behaves linearly with
$E^2$:  $J(E^2)\sim \alpha^\prime E^2
z_{IR}^2/R^2$. That is, the 
AdS open string for finite $z_{IR}$ behaves 
like a massless relativistic string
at low energy \cite{goddardgrt}. The effective tension of this
slightly stretched string is $T_{\rm eff}=T_0R^2/z_{IR}^2$ 
and drops to zero when the 
IR cutoff is removed, $z_{IR}\to\infty$, causing the
mass spectrum to become continuous from zero.
At higher energy the energy spectrum shows a
threshold $E_I=R^2/(\pi\alpha^\prime z_{IR})$ 
at which there is an accumulation of
energy levels. $J$ becomes infinite as $E\to E_I$ from
below, in a manner similar to the Bohr levels in hydrogen just below
ionization threshold. Specifically, $E(J)\sim E_I(1-
\alpha^2/(8J^2))$, where $\alpha=4\pi^2R^2/(\alpha^\prime\Gamma(1/4)^4)$. 
This behavior at large $J$ agrees with the nonrelativistic model
of two particles with equal mass, given by the energy
of a static string stretching from $z=z_{IR}$ to $z=\infty$, 
interacting with a 
Coulomb potential $-\alpha/r$. Here $\alpha$ is precisely
equal to the value predicted
by the AdS/CFT correspondence for the $Q{\bar Q}$ static force  
in ${\cal N}=4$ Yang-Mills theory in the strong 't Hooft coupling
limit \cite{maldacenaqqbar}. 
\section{Covariant four open string scattering in flat space}
We begin our study of classical worldsheets 
with a translation of known results for scattering of strings
in flat Minkowski space to the T-dual description:
\bea
{\partial y^\mu\over\partial x}=i{\partial x^\mu\over \partial y},\qquad
{\partial y^\mu\over\partial y}=-i{\partial x^\mu\over \partial x},
\eea
where we parameterize the worldsheet using the upper half complex plane
$z=x+iy$. Then for $n$-string scattering we have
\bea
x^\mu(z)&=&{-i}\sum_{r=1}^{n-1}{(y_r-y_{r-1})^\mu\over\pi}\log|z-u_r|,\\
y^\mu(z)&=&\sum_{r=0}^{n-1}{y_r^\mu\over\pi}\left(\tan^{-1}{u_{r+1}-x\over y}
-\tan^{-1}{u_r-x\over y}\right).
\eea
Here the momenta carried by the strings are $p_r=y_r-y_{r-1}$, and the
Koba-Nielsen variable of string $r$ is $u_r$. For definiteness we
take $0=u_1<u_2<\cdots<u_{n-1}=1$ and $u_0=-\infty$, $u_n=\infty$.
The $x^\mu$ satisfy Neumann conditions $\partial x^\mu/\partial y =0$
on the real axis $y=0$ (away from the points
$x= u_r$), whereas the $y^\mu$ satisfy Dirichlet conditions on the
real axis: $y^\mu(x,0)= y_r^\mu$ on the interval $u_r<x<u_{r+1}$.
In T-dual variables the $y^\mu$ are discontinuous where the
string momenta enter the worldsheet. Specializing to $n=4$:
\bea
x^\mu(z)&=&{-i}\sum_{r=1}^{3}{(y_r-y_{r-1})^\mu\over\pi}\log|z-u_r|,\\
y^\mu(z)&=&{y_0^\mu\over\pi}\left({\pi\over2}-\tan^{-1}{x\over y}\right)
+{y_1^\mu\over\pi}\left(\tan^{-1}{u-x\over y}+\tan^{-1}{x\over y}\right)\nonumber\\
&&+{y_2^\mu\over\pi}\left(\tan^{-1}{1-x\over y}-\tan^{-1}{u-x\over y}\right)+
{y_3^\mu\over\pi}\left({\pi\over2}-\tan^{-1}{1-x\over y}\right).
\eea
Strictly speaking, conformal gauge also requires the constraints
$(\partial x^\mu/\partial x)(\partial x_\mu/\partial x)=(\partial x^\mu/\partial y)(\partial x_\mu/\partial y)$ and $(\partial x^\mu/\partial x)(\partial x_\mu/\partial y)=0$. These would
be expected to hold only on-shell though. For the classical
string on-shell means $(y_{i+1}-y_i)^2=0$. But even then
these constraints only hold when the Koba-Nielsen variables are at a
stationary point of the action, which corresponds to a saddle point
of the integration. (For example 
$u=u_0\equiv t/(s+t)$ for the four point function). 
Away from the saddle point the $u$ integrand is off-shell in the
classical sense. Integration over the whole range
puts the amplitude on-shell in
the quantum sense. Indeed we know that the integrated Koba-Nielsen
integrands do satisfy the quantum operator Virasoro constraints. 
It is interesting that the classical Virasoro constraints do hold,
though, at the saddle points. 

\begin{figure}[ht]
\begin{center}
\includegraphics[width=3in,height=3in]{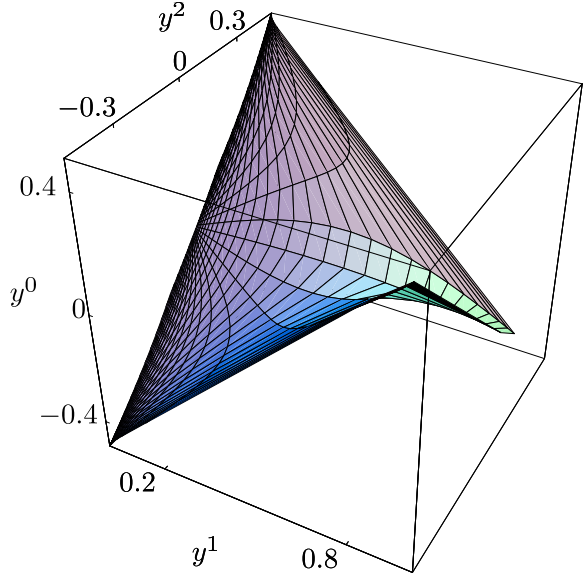}
\caption{The worldsheet in dual $y$ space for the scattering solution of
a string in flat space. The kinematics is $s,t <0$.
The case shown is
$u=0.5$ or $s=t$.}
\label{flat4pt}
\end{center}
\end{figure}

An interesting feature of the
T-dual worldsheet is that it is finite in extent. We display
this graphically in Fig.~\ref{flat4pt}. It is striking that, plotted
in $y$ space, the boundary
of the worldsheet is a finite polygon of lightlike straight line segments.
It is important to realize that in the conformally flat coordinates
$x,y$, these segments are the images of points on the $x$ axis---the
points where the boundary values of $y$ are discontinuous. Since in the
$x,y$ coordinates the problem is just a Dirichlet problem
for the Laplace equation, the solution is uniquely determined
by the values of $y$ specified on the boundary, even if there
are discontinuities in these boundary values. In Fig.~\ref{flat4pt}
the Dirichlet data are just the four corners of the polygon. The
reason that the discontinuities in $y^\mu$ are mapped to straight lines
in this figure is that, in the upper half plane 
near a discontinuity at say $x+iy=u_r$, the different components
of $y^\mu$ are related to their boundary values by the same Green
functions and hence are linearly related to each other. Thus, 
as $u_r$ is approached from different directions in the upper half plane,
the solution $y^\mu(x,y)$ approaches different points on a straight
line joining the two boundary values on either side of $u_r$. 
It might seem strange that giving data at 4 points is enough to determine
the solution, but we must remember that if we parametrize the
worldsheet with coordinates in $y^\mu$-space (e.g. $y^1,y^2$), the equations
for the worldsheet are nonlinear and quite different from the Laplace
equation.
\section{The string worldsheet on AdS$_3$}
Next we turn to a consideration of the worldsheet solution
found by AM which controls gluon scattering at strong
coupling in ${\cal N}=4$ super Yang-Mills theory.
We will look in more detail at their method and find 
additional solutions whose interpretation is quite different.

Starting from the Nambu-Goto action (\ref {ynambugoto}) in terms of T-dual 
variables which we call $y'^\mu,\; r'$,\footnote{In this section the prime no longer indicates a derivative.} we restrict ourselves to AdS$_3$
and make the Anzatz, closely related to one proposed earlier
by Kruczenski \cite{kruczenski}, and used by AM:
\bea r'=w(\xi^0)e^{\xi^0},\quad y'^0=e^{\xi^0}\cosh{\xi^1},
\quad y'^1=e^{\xi^0}\sinh\xi^1
,\quad y'^2=y'^3=0.\eea 
The Lagrangian whose stationary point determines $w(\xi^0)$ is then 
\bea L=-{\sqrt{1-(w+\dot w)^2}\over w^2}\eea 
and the ensuing Euler-Lagrange equation is
\bea -{d\over d \xi^0}{w+\dot w\over w^2\sqrt{1-(w+\dot w)^2}}=
-{w+\dot w\over w^2\sqrt{1-(w+\dot w)^2}}-2{\sqrt{1-(w+\dot w)^2}\over w^3}\eea 
where the dot indicates derivative with respect to $\xi^0$.

Note that for a static solution, $\dot w=0$, 
we immediately find $w=\pm\sqrt{2}$.  
For positive $r'$ one would take $w=\sqrt{2}$ which is 
the starting point of AM.  To find nonstatic solutions 
we can use the conservation law that follows from $L$ to find
\bea E={1-w(w+\dot w)\over w^2\sqrt{1-(w+\dot w)^2}},\eea 
\bea \dot w={1-w^2-E^2w^4\pm|w|\sqrt{E^4w^4+(w^2-1)E^2}\over
w(1+E^2w^2)},\eea 
\bea {d\xi^0\over d|w|}={|w|\over 1-w^2}
\left[1\pm{|w|\over\sqrt{w^4+{(w^2-1)/E^2}}}\right].\label {sqrt}.\eea 
The condition that 
$({\partial y'/\partial \xi^1})\cdot
({\partial y'/\partial \xi^1})=0$ at $r'=0$  
requires that the region $\xi^0\rightarrow-\infty$ be part of the solution.  
For generic $E^2$ the singularities arising from the zeros of the square root 
are integrable, thereby leaving only $|w|\rightarrow1$ and $|w|\rightarrow\infty$ 
as possibilities.  At large  $|w|$ the integrand behaves as $-d|w|/|w|$ 
implying $\xi^0\sim -\log|w|$ or $|w|\sim e^{-\xi^0}$.   
But then $r'\sim{\rm O}(1)$.  This is not acceptable since,  for the 
application to gluon scattering, we need the corner to be at $r'=0$.

We next examine the behavior near $|w|=1$ which is 
singular only for the upper sign in (\ref {sqrt}):
\bea d\xi^0\sim - d|w| {1\over|w|-1}=-d\log\left||w|-1|\right|.\eea 
This singularity corresponds to $\xi^0\rightarrow+\infty$.  
We conclude, therefore, that the solution for generic $E^2$  
does not include the light-like corner at $r'=0$.

\begin{figure}[ht]
\begin{center}
\includegraphics[width=3in]{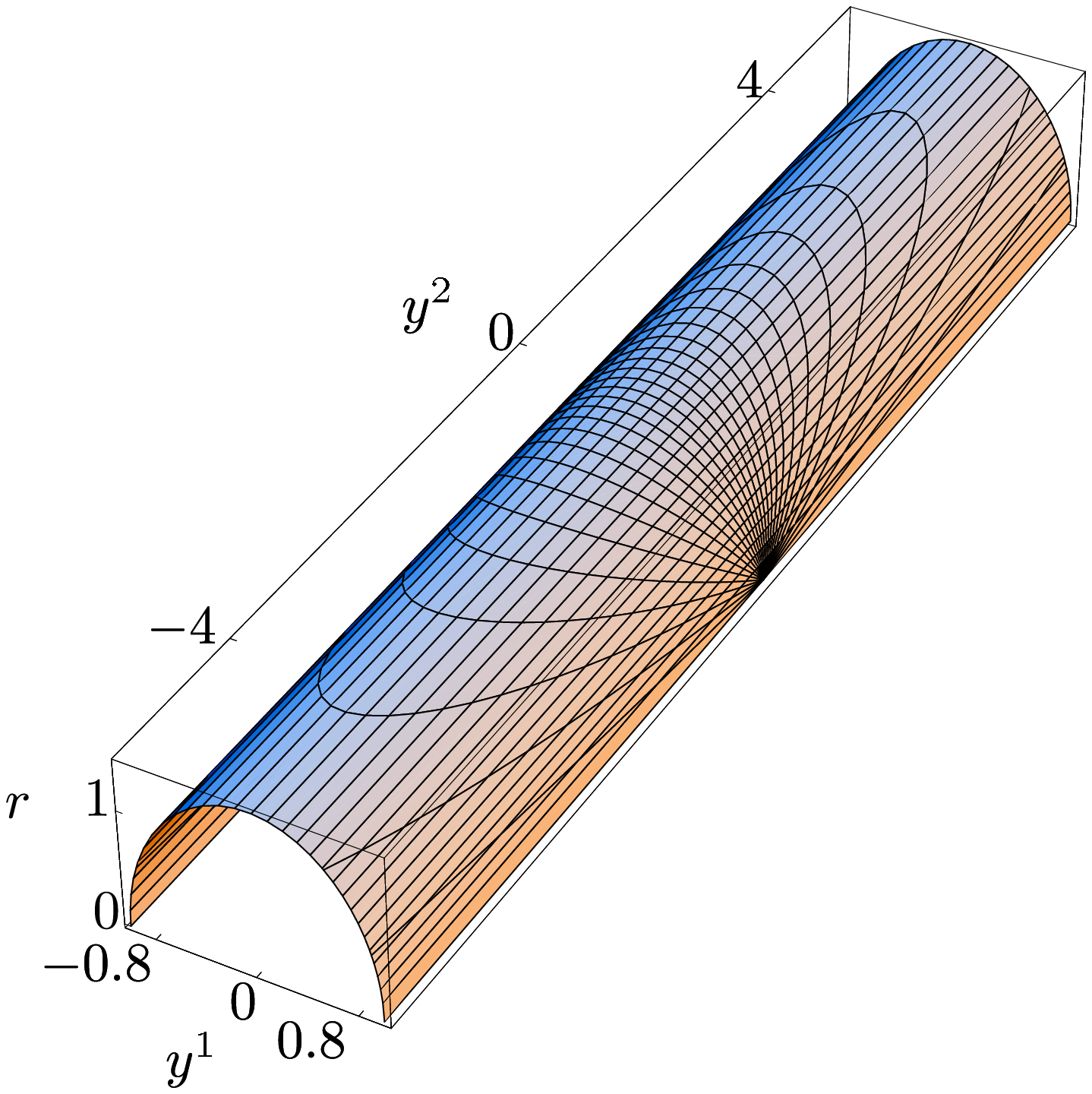}\quad
\includegraphics[width=2in]{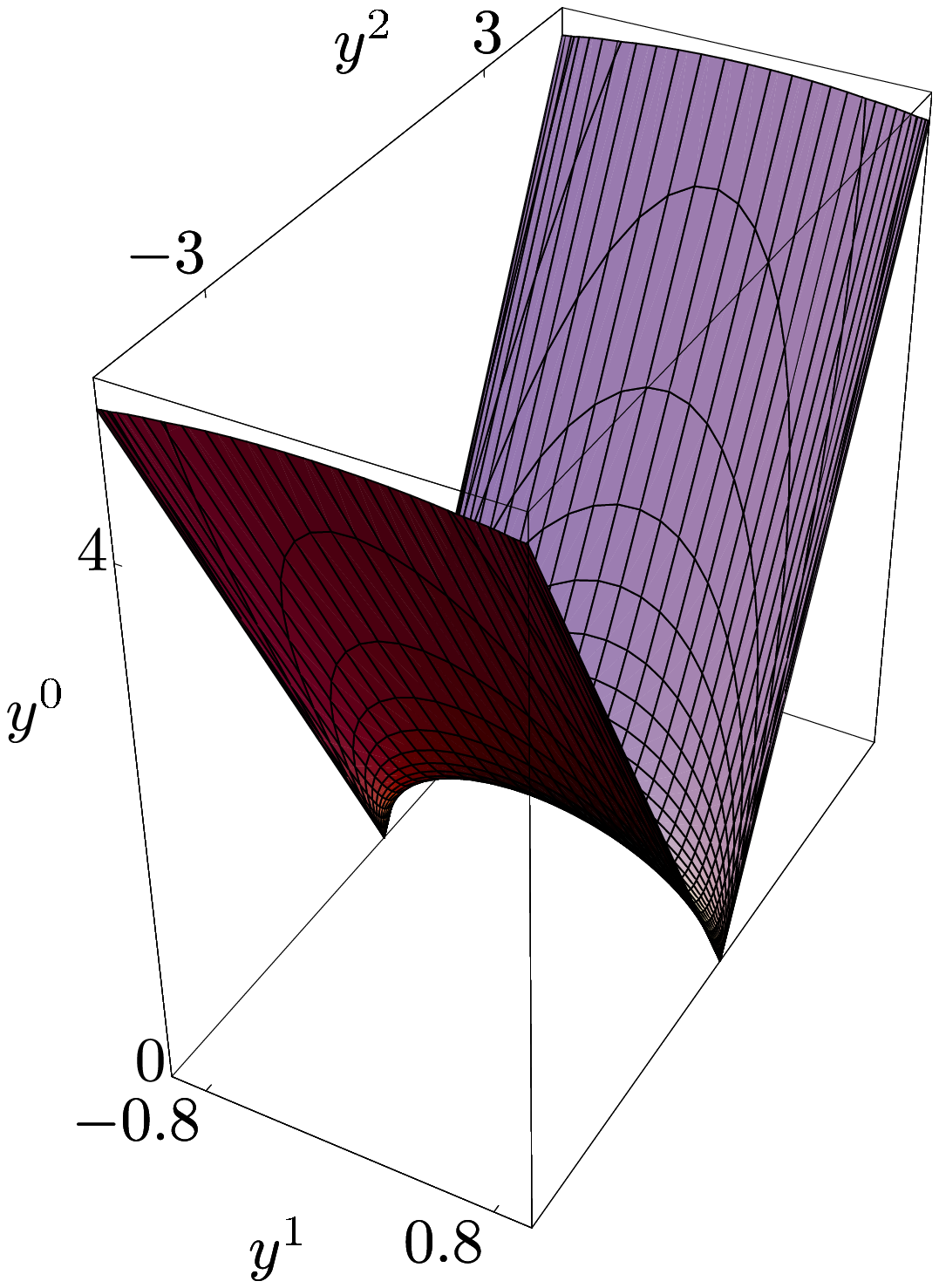}
\caption{$r$ and $y^0$ plotted against $y^1$ and $y^2$ for $\theta=0$.}
\label{adsws4glue1}
\end{center}
\end{figure}

\begin{figure}[ht]
\includegraphics[width=3in]{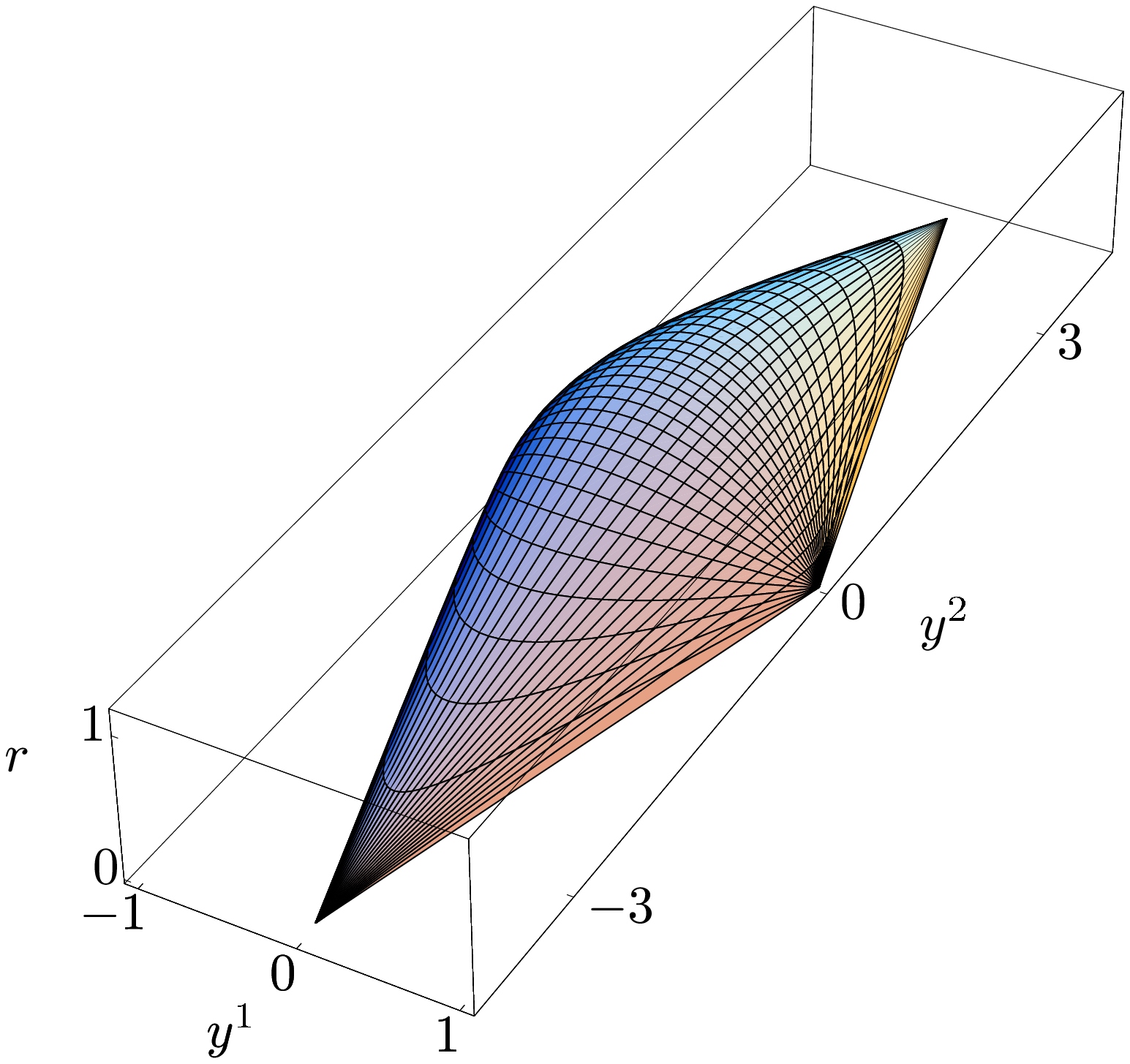}$\quad$
\includegraphics[width=2.5in]{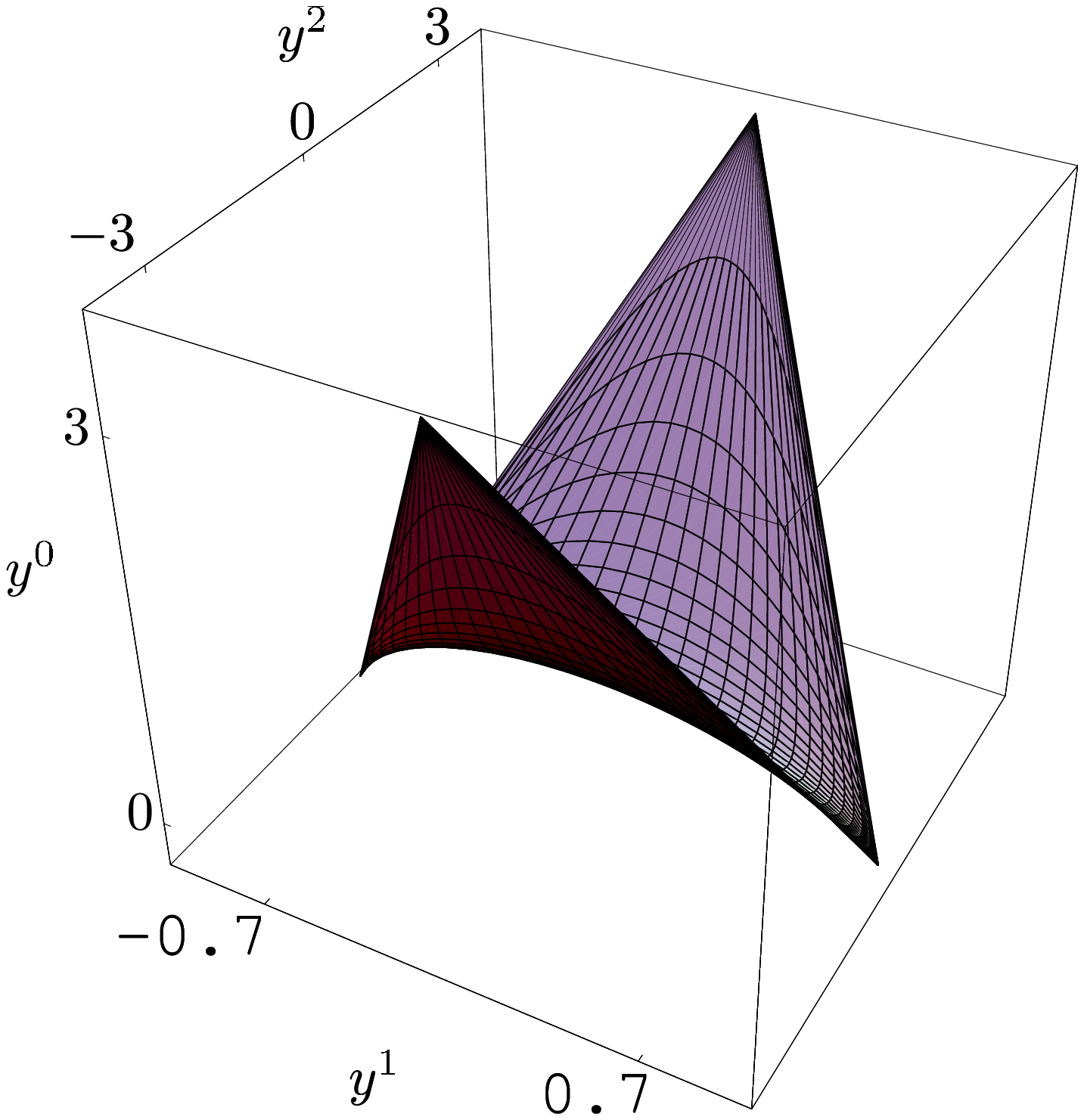}
\caption{$r$ and $y^0$ plotted against $y^1$ and $y^2$ for $\theta=\pi/12$.}
\label{adsws4glue2}
\end{figure}

\begin{figure}[ht]
\includegraphics[width=3in]{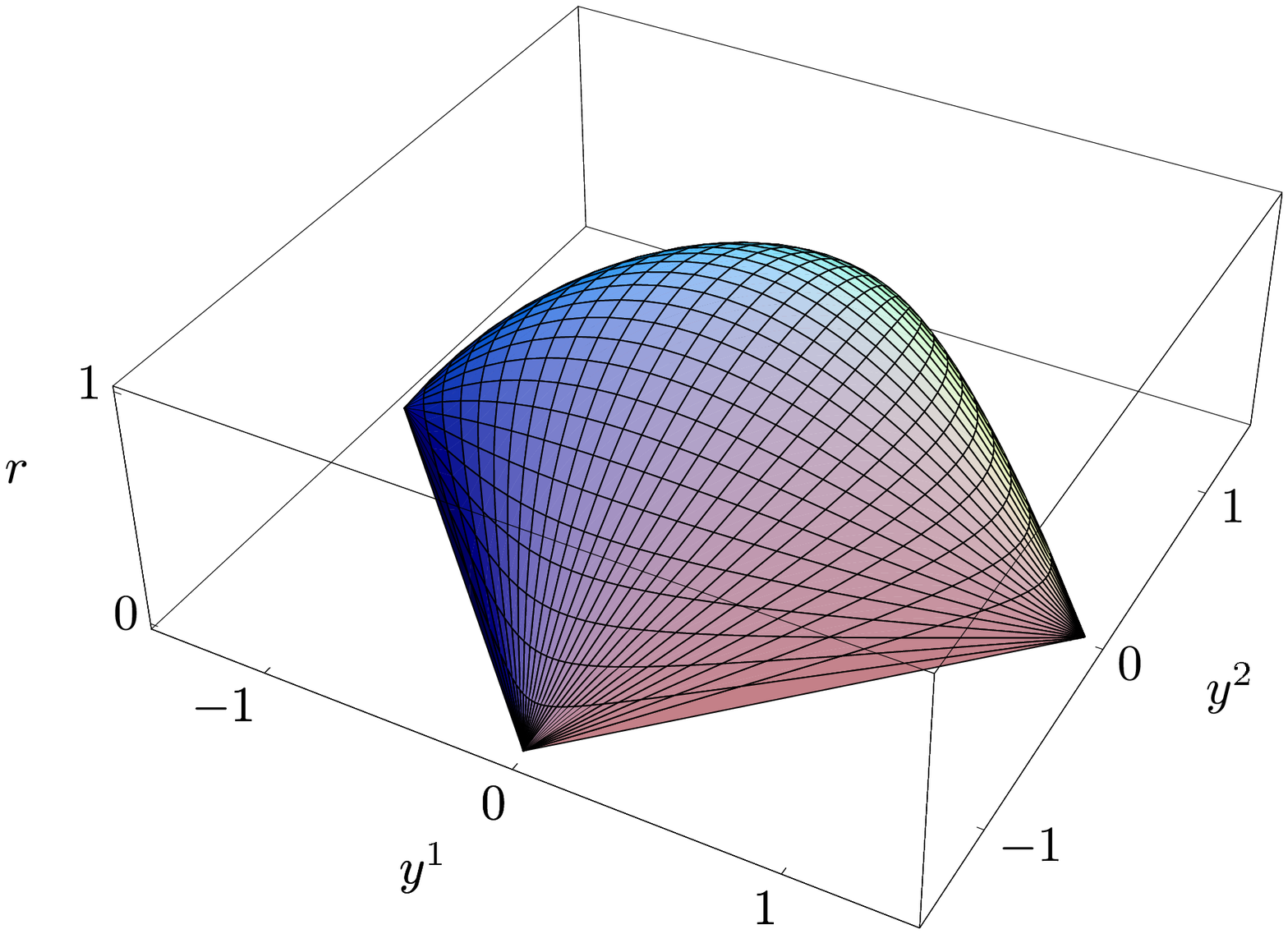}$\quad$
\includegraphics[width=2.5in]{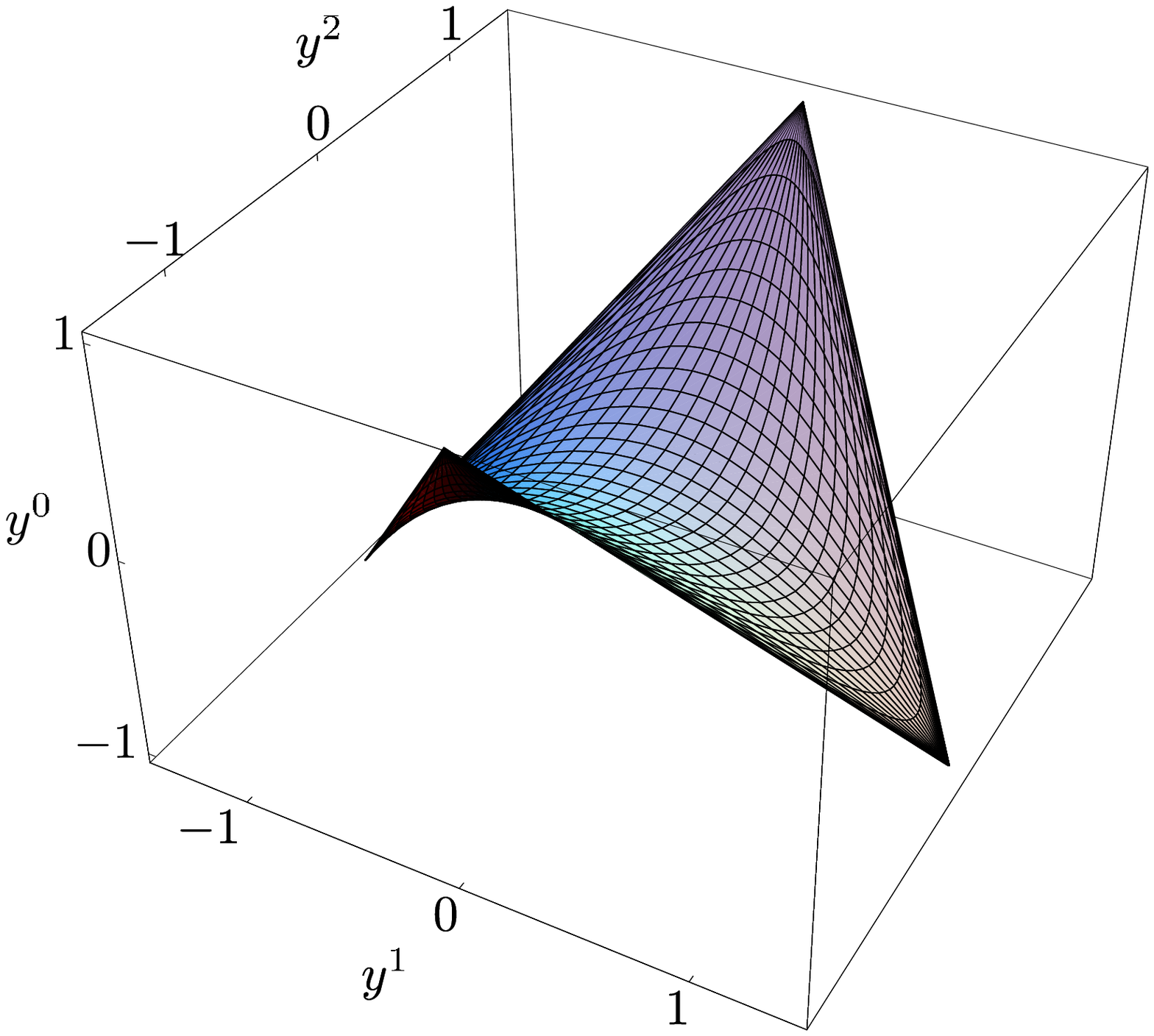}
\caption{$r$ and $y^0$ plotted against $y^1$ and $y^2$ for $\theta=\pi/4$.}
\label{adsws4glue3}
\end{figure}

For the special case $E^2=-1/4$, however, 
the zeros of the square root in (\ref{sqrt}) coalesce and the 
singularity becomes a pole at $|w|=\sqrt{2}$.  This 
corresponds to the desired behavior $\xi^0\rightarrow -\infty$.  
Then (\ref{sqrt}) becomes
\bea {d\xi^0\over d|w|}=-{|w|(2-|w|)\over(2-w^2)(|w|-1)}\eea 
which integrates to
\bea C e^{\xi^0}={1\over|w|-1}
\left(\sqrt2-|w|\over\sqrt2+|w|\right)^{1/\sqrt2}\equiv f(|w|)\eea 
where $C$ is an integration constant.  
Note that as $C\rightarrow0$ we obtain the static 
solution $|w|=\sqrt{2}$.  We get a completely different solution 
by choosing $C=1$ and taking $w>0$ in order that $r'\ge0$.  
Both solutions have $E^2=-1/4$.  A plot of $w(\xi^0)$ for the 
nonstatic solution is given in Fig.~\ref{nonstatic}.
{\begin{figure}[ht]
\begin{center}
\psfrag{x}{$\xi^0$}
\psfrag{y}{$w$}
\includegraphics[width=3in]{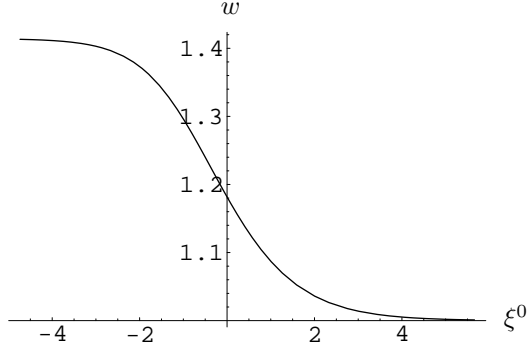}
\caption{$w$  plotted against $\xi^0$ }
\label{nonstatic}
\end{center}
\end{figure} 
Note that  $w(\xi^0)$ approaches the static 
solution $\sqrt{2}$ as $\xi^0\rightarrow-\infty$.

It will prove convenient in what follows to replace $\xi^0$ as one of the worldsheet parameters by $w$.  Then we can write both the static and nonstatic solutions using a common notation.  We have
\bea y'^0=f(w){\cosh\xi^1}, \quad y'^1=f(w)\sinh\xi^1, \quad r'=g(w)f(w)\label{ansatz}\eea 
where $g(w)=\sqrt2$ for the static solution and $g(w)=w$ for the nonstatic solution.

\section{The four corner solution on AdS$_5$}

In the previous section we described a worldsheet
solution for the dual variables that lies in an AdS$_3$ subspace of 
AdS$_5$ by assuming (\ref {ansatz}) along with $y'^2=y'^3=0$.  
Any such solution can be extended into AdS$_5$ by means of SO(4,2) 
rotations and boosts of the embedding coordinates 
\bea Y'^\mu={R\over r'} y'^\mu ,\quad Y'^{-1}
={1\over2r'}(R^2+r'^2+y'^\mu y'_\mu),\quad Y'^4
={1\over 2r'}(R^2-r'^2-y'^\mu y'_\mu)\eea 
which are such that
\bea (Y'^{-1})^2-(Y'^4)^2-Y'^\mu Y'_\mu=R^2.\eea 
Notice that $Y^{\prime4}$ and $Y^{\prime-1}$ enter the left side 
just as space-like and time-like coordinates respectively.
Thus we are free to reinterpret the AdS$_3$ with the relabeling
$4\to1\to2\to4$ so that $Y'^4=0$ in the new interpretation.
Then if we make rotations given by the angle $\theta$ 
in the (0,-1) plane and $\phi$ in the new (1,2) plane 
as well as a boost through hyperbolic angle $\alpha$ 
in the  (0,4) plane, we obtain a new dual AdS$_5$ solution:
\bea r=R{2r'R\over (R^2+r'^2+y'^\mu y'_\mu)
(\cos\theta+\sin\theta\sinh\alpha)
+2y'^0R(\sin\theta-\cos\theta\sinh\alpha)}\eea 
\bea y^0=R{\cosh\alpha[(2 y'^0R\cos \theta
-(R^2+r'^2+y'^\mu y'_\mu)\sin\theta]\over
(R^2+r'^2+y'^\mu y'_\mu)(\cos\theta+\sin\theta\sinh\alpha)
+2y'^0R(\sin\theta-\cos\theta\sinh\alpha)}\eea 
\bea y^1=R{(R^2-r'^2-y'^\mu y'_\mu)\cos\phi+2y'^1R\sin\phi\over
(R^2+r'^2+y'^\mu y'_\mu)(\cos\theta+\sin\theta\sinh\alpha)
+2y'^0R(\sin\theta-\cos\theta\sinh\alpha)}\eea 
\bea y^2=R{2 y'^1 R\cos\phi -(R^2-r'^2-y'^\mu y'_\mu)\sin\phi\over
(R^2+r'^2+y'^\mu y'_\mu)(\cos\theta+\sin\theta\sinh\alpha)
+2y'^0R(\sin\theta-\cos\theta\sinh\alpha)}\eea 
\bea y^3=0.\eea 
More precisely, what we are doing here is viewing the original
solution from the point of view of different frames as well as
different Poincar\'e patches.

Ryang \cite{ryang} 
has conducted an investigation of the behavior of the 
solution employed by AM under more general SO(4,2) transformations.  
AM take $\theta=\phi=\pi/4$ and $\sinh \alpha=v$.    
In the following, we restrict attention to $\phi=\alpha=0$.
Using the AdS$_3$ Ansatz (\ref{ansatz})
\bea 
r'=g(w)f(w),\quad y'^0=f(w)\cosh{\xi^1},\quad y'^1=f(w)\sinh\xi^1
,\quad y'^2=y'^3=0
\eea 
we find
\bea r=R{g(w)\over c(w) \cos\theta+\cosh\xi^1\sin\theta}\label{firsteq}\eea 
\bea y^0=R{\cosh\xi^1\cos\theta-c(w)\sin\theta\over c(w) \cos\theta+\cosh\xi^1\sin\theta}\eea 
\bea y^1=R{-s(w)\over c(w)\cos\theta+\cosh\xi^1\sin\theta}\eea 
\bea y^2=R{\sinh\xi^1\over c(w)\cos\theta+\cosh\xi^1\sin\theta}\eea 
\bea y^3=0\label{lasteq}\eea 
where 
\bea c(w)={1\over2}\left(f(w)[\{g(w)\}^2-1]+{1\over f(w)}\right),\quad s(w)={1\over2}\left(f(w)[\{g(w)\}^2-1]-{1\over f(w)}\right),\eea 
We assume  that $0\le\theta\le\pi/2$ so as to avoid $r<0$ and find, for $g(w)=\sqrt2$,
\bea  r=R{\sqrt{2}\over \cosh \xi^0 \cos\theta+\cosh\xi^1\sin\theta}
\label{4r}\eea
\bea y^0=R{\cosh\xi^1\cos\theta-\cosh\xi^0\sin\theta\over \cosh \xi^0 \cos\theta+\cosh\xi^1\sin\theta}
\label{4y0}\eea
\bea y^1=R{-\sinh\xi^0\over\cosh \xi^0 \cos\theta+\cosh\xi^1\sin\theta}
\label{4y1}\eea
\bea y^2=R{\sinh\xi^1\over\cosh \xi^0 \cos\theta+\cosh\xi^1\sin\theta}
\label{4y2}\eea
\bea y^3=0.
\label{4y3}
\eea
We illustrate this solution in 
Figs.\ \ref{adsws4glue1}-\ref{adsws4glue3}} for several angles $\theta$.   
Note that in this case $c(w)=\cosh\xi^0$ and $s(w)=\sinh\xi^0$.
For $\pi/4\le\theta\le\pi/2$ we can make use of the symmetry 
under $\theta\leftrightarrow\pi-\theta$, $\xi^0\leftrightarrow\xi^1$, 
$y^1\leftrightarrow -y^2$ and $y^0\leftrightarrow-y^0$. Notice that
when $\theta>0$ these solutions show four corners, but when $\theta=0$
a pair of these corners is sent to infinity, leaving only two corners.

The physical meaning of a solution  is 
found by looking in the Minkowski target space of $y^\mu$ at points
 where $r=0$ (or almost 0).  Using (\ref{4r}) we find
that $r=0$ when $|\xi^0|$ or $|\xi^1|$ or possibly both are infinite.  
There are four cases, where one or the other of 
$|\xi^0|$ and $|\xi^1|$ are infinite, with the other finite:
\bea \xi^0\rightarrow\infty:\quad y^\mu_a=R(-\tan\theta,-\sec\theta,0,0)\eea 
\bea \xi^1\rightarrow  \infty:\quad y^\mu_b=R(\cot\theta,0,\csc\theta,0)\eea 
\bea \xi^0\rightarrow -\infty:\quad y^\mu_c=R(-\tan\theta,\sec\theta,0,0)\eea 
\bea \xi^1\rightarrow  -\infty:\quad y^\mu_d
=R(\cot\theta,0,- \csc\theta,0).\eea 
In the interpretation of these solutions as describing scattering of
glue by glue, the differences of these quantities are proportional to the 
(incoming) momenta of the massless gluons.
We should then have
\bea (y_a-y_b)^2=(y_b-y_c)^2=(y_c-y_d)^2=(y_d-y_a)^2
=R^2[\sec^2\theta+\csc^2\theta-(\tan\theta+\cot\theta)^2]=0\eea 
which is indeed the case.  Furthermore, we can identify the 
Mandelstam variables for the gluon scattering process as 
\bea s=-(y_a-y_c)^2=-4R^2\sec^2\theta\eea 
\bea t=-(y_b-y_d)^2=-4R^2\csc^2\theta\eea 
so that $t/s=\cot^2\theta$.

Using the duality transformation (\ref{genduality}) we can translate
this solution to the original coordinate space, with $z=R^2/r$
in place of $r$:
\bea  z={R\over\sqrt{2}}(\cosh\xi^0\cos\theta+\cosh\xi^1\sin\theta)\eea 
\bea  x^0=i{R\over2}\sinh\xi^0\sinh\xi^1\eea 
\bea  x^1=i{R\over2}( \xi^1\cos\theta+\cosh\xi^0\sinh\xi^1\sin\theta)\eea 
\bea  x^2=i{R\over2}(\xi^0\sin\theta+\sinh\xi^0\cosh\xi^1\cos\theta)\eea 
\bea  x^3=0.\eea 
Note the world sheet invariance under 
$\theta\rightarrow\pi/2-\theta$, $\xi^0\leftrightarrow\xi^1$, 
$y^1\leftrightarrow -y^2$, $x^1\leftrightarrow x^2$. The fact that
the coordinates are all imaginary shows that the physical 
scattering process
described is classically inaccessible, analogous to reflection
above a barrier in one-dimensional quantum mechanics.
\section{The one-corner solution for AdS$_5$}
\begin{figure}[ht]
\includegraphics[width=3in]{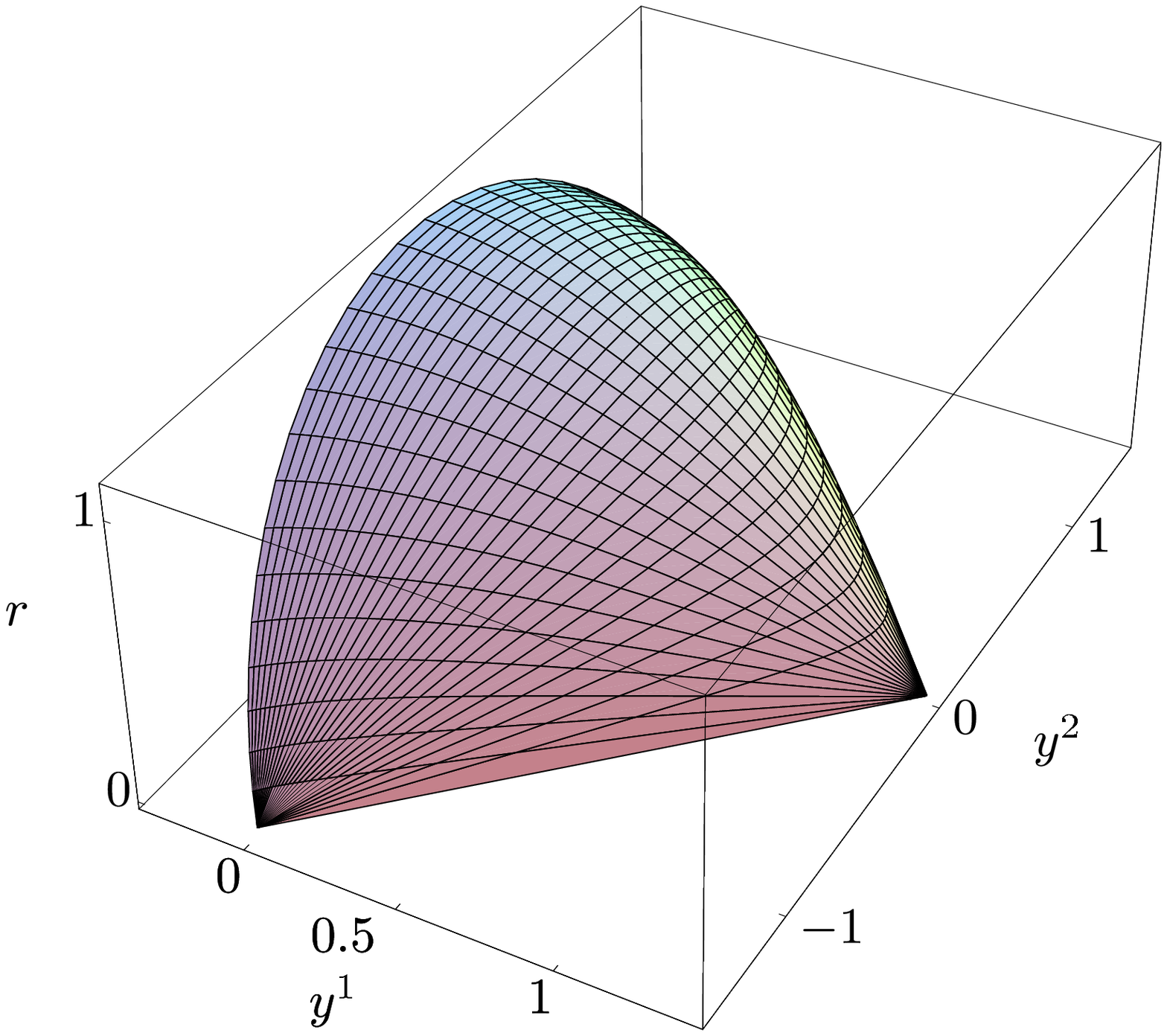}
\includegraphics[width=3in]{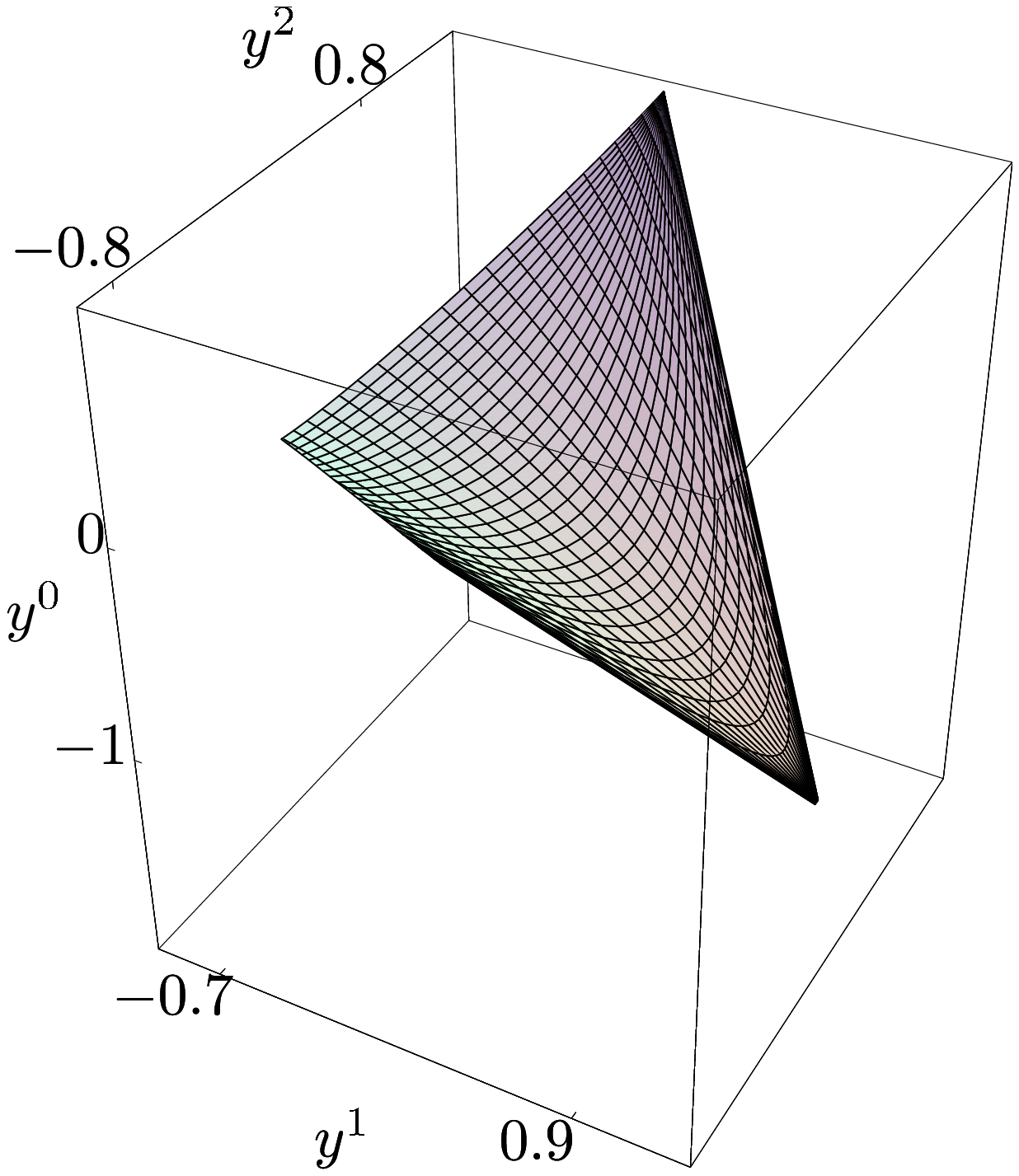}
\caption{$\bar r$ and $\bar y^0$ plotted against 
$\bar y^1$ and $\bar y^2$ for $\theta=\pi/4$, as $w$ runs from 1 to $\sqrt2$.}
\label{adsws4glue8}
\end{figure}

\begin{figure}[ht]
\includegraphics[width=3in]{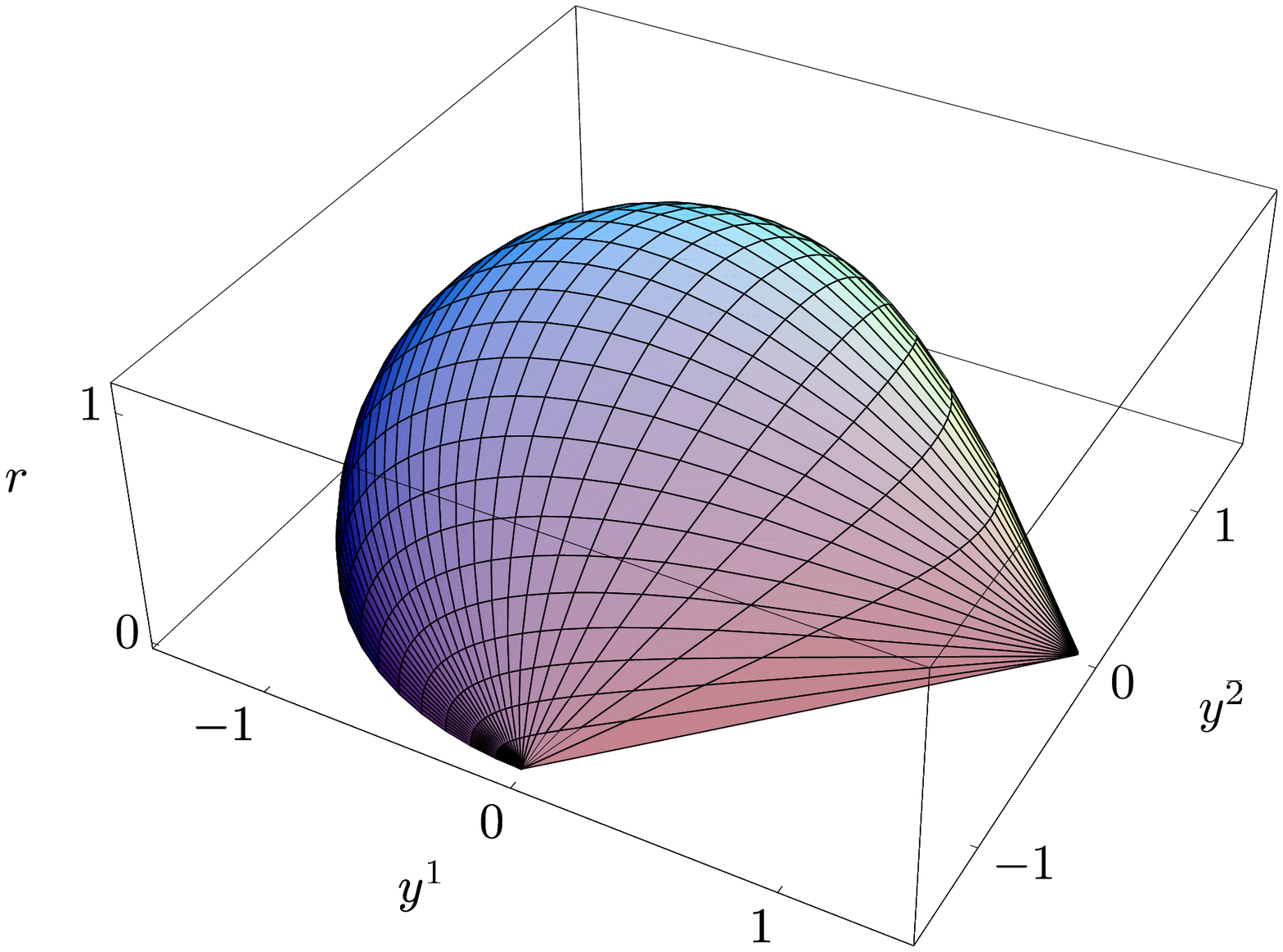}
\includegraphics[width=3in]{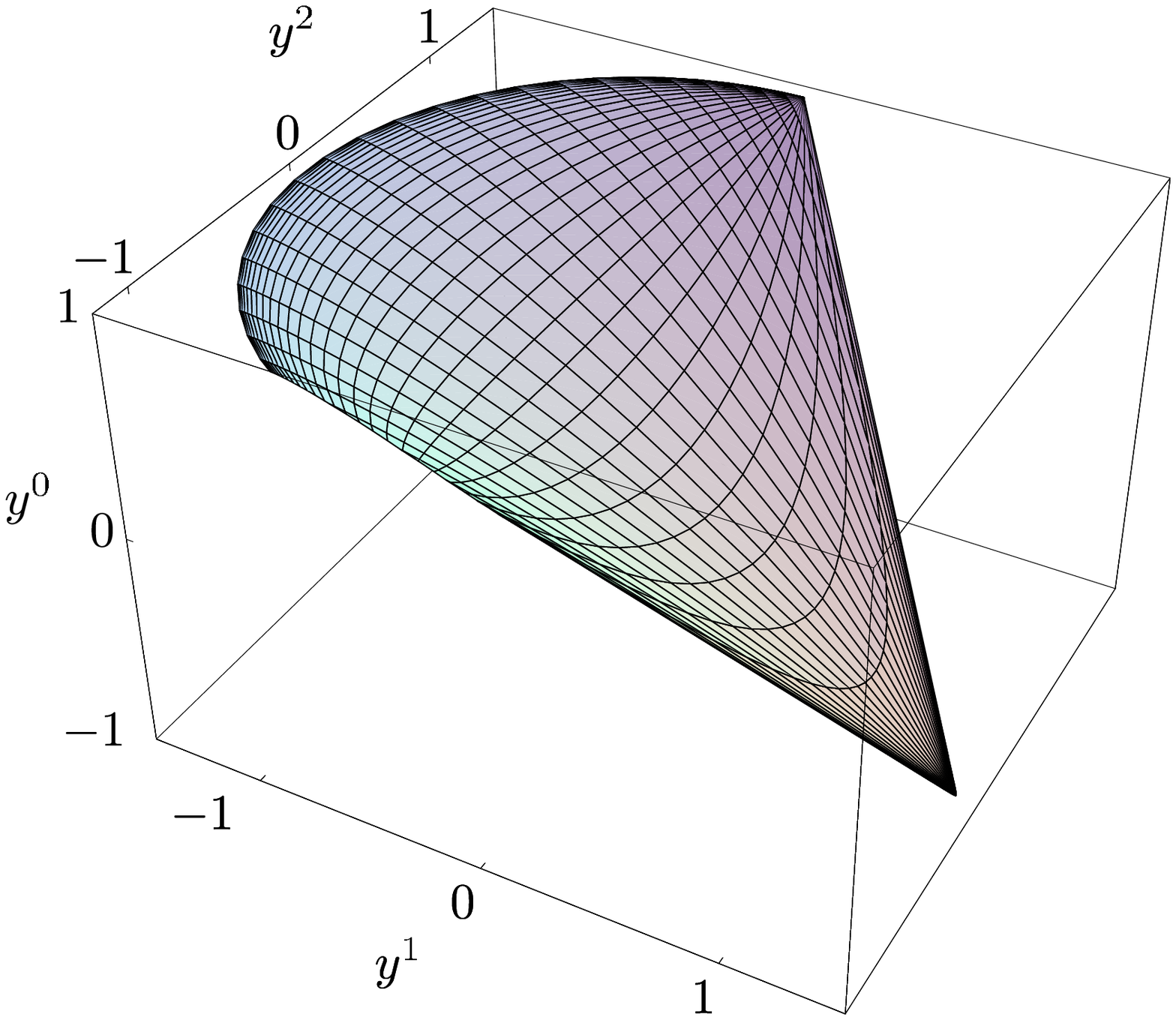}
\caption{$\bar r$ and $\bar y^0$ plotted against 
$\bar y^1$ and $\bar y^2$ for $\theta=\pi/4$, as $w$ runs from 0 to $\sqrt2$.}
\label{adsws4glue11}
\end{figure}

We next examine the properties that follow 
from using the nonstatic solution in the embedding into AdS$_5$.  
This is accomplished by choosing $g(w)=w$.  
This new solution is given by Eqs.\ (\ref{firsteq})-(\ref{lasteq})
\bea \bar r=R{w\over c(w) \cos\theta+\cosh\xi^1\sin\theta}\eea 
\bea \bar y^0=R{\cosh\xi^1\cos\theta-c(w)\sin\theta\over c(w) 
\cos\theta+\cosh\xi^1\sin\theta}\eea 
\bea \bar y^1=R{-s(w)\over c(w)\cos\theta+\cosh\xi^1\sin\theta}\eea 
\bea\bar y^2=R{\sinh\xi^1\over c(w)\cos\theta+\cosh\xi^1\sin\theta'}\eea 
\bea \bar y^3=0.\eea 
We illustrate the result in Fig.~\ref{adsws4glue8} for $\theta=\pi/4$. 
What we see are two light-like lines in the $r=0$ hyperplane 
joined at one corner.  They are connected at their other ends 
by a curve for which $r\ne0$.   However, having made the SO(4,2) 
transformation it is now meaningful to extend the domain of $w$ 
to the region $w<1$ where $r$ remains positive.\footnote{We thank
Martin Kruczenski for stressing this possibility to us.}  
If we do this 
Fig.~\ref{adsws4glue8} now is transformed to Fig.~\ref{adsws4glue11}.

We are once again interested in what happens when $\bar r=0$.  
This occurs when $w\rightarrow0$ (assuming $\theta>0$), 
when $w\rightarrow\sqrt2$ 
and when $\xi^1\rightarrow\pm\infty$.   We now have
\bea w\rightarrow0:\quad \bar y^\mu_a=R(\cot\theta,
-\csc\theta\,{\rm sech }\,\xi^1,\csc\theta\tanh\xi^1,0)\label{curve}\eea 
\bea \xi^1\rightarrow  \infty:\quad \bar y^\mu_b
=R(\cot\theta,0,\csc\theta,0)\label{pointa}\eea 
\bea w\rightarrow \sqrt2:\quad \bar y^\mu_c=R(-\tan\theta,\sec\theta,0,0)\eea 
\bea \xi^1\rightarrow  
-\infty:\quad \bar y^\mu_d=R(-\tan\theta,0,- \csc\theta,0).\label{pointd}\eea 
The curve described by (\ref{curve}) is a 
semi-circle of radius $\csc\theta$ lying in the plane 
given by ${\bar r}=0$ and constant $\bar y^0=\cot\theta$ that 
runs between the points $\bar  y^\mu_b$ and $\bar y^\mu_d$.  
The two lines that run from $\bar y^\mu_c$ to $\bar y^\mu_b$ 
and from $\bar y^\mu_c$ to $\bar y^\mu_d$ are lightlike and 
lie in the plane given by ${\bar r}=0$ and $\bar y^0\sin\theta+\bar y^1=\cos\theta$.
The two light-like line segments can be identified with the
incoming momenta of two gluons involved in a scattering
process off an external source. The process is described in terms
of the dual coordinates $\bar y^\mu$ of a string, one end of which
stays fixed at ${\bar y}_c$ (corresponding to a free end in $\bar x$ space)
and the other end of which follows the semi-circle. 
The fact that this semi-circular part of the boundary is at constant $\bar y^0$
means that the source conserves energy throughout the scattering
process, in other words the source is static.

To find the coordinate-space representation of this solution we calculate 
\bea {\bar r}^\prime{\dot {\bar r}}+\bar y'\cdot{\dot{\bar y}}=0,\quad 
{{\dot {\bar y}}^2+{\dot {\bar r}}^2 \over \bar y'^2+\bar r'^2}=
{4{\dot w}^2\over(2-w^2)^2}
\eea
which yields
\bea \mu=0,\quad \kappa=-2i{w-1\over w(2-w)}.\eea
Then the duality conditions (\ref{genduality}) read
\bea {\partial{\bar x}^\mu\over\partial\xi^1}
={R^2\over \kappa {\bar r}^2}{\partial\bar y^\mu\over\partial \xi^0}
=-i{R^2\over {\bar r}^2}{ 2-w^2\over2}{\partial\bar y^\mu\over\partial w}
\eea
\bea{\partial{\bar x}^\mu\over\partial w}
={R^2\kappa\over  {\bar r}^2}{\partial\bar y^\mu\over\partial\xi^1}
{\partial \xi^0 \over\partial w}=i{R^2\over{\bar r}^2}{2\over2-w^2}
{\partial\bar y^\mu\over\partial\xi^1}
\eea
which leads to
\bea \bar x^0=iR{(2-w^2)\over 2w^2}c'(w)\sinh \xi^1\eea
\bea\bar x^1=iR{2-w^2\over 2w^2}
\left(s^\prime(w)\sinh\xi^1\sin\theta-
\xi^1\cos\theta{w^2\over 2-w^2}\right)\eea
\bea\bar x^2=iR{2-w^2\over 2w^2}c'(w)\cosh\xi^1\cos\theta 
-iR\sin\theta\left({1\over w}+{1\over 2\sqrt2}
\log{\sqrt2-w\over\sqrt2+w}\right),\eea
to which we can add real parts that are independent of $\xi^1$ and $w$.  These coordinates are concentrated at a real spacetime point.  As chosen the worldsheet in coordinate space is entirely imaginary and therefore classically inaccessible as in reflection above a barrier.  As a consequence the scattering
amplitude corresponding to this solution is exponentially suppressed. 
We also give the $w\to0$ boundary curve in coordinate space:
\bea
{{\bar x}^{0}} &\sim&{iR\over2}\sinh\xi^1\\
{{\bar x}^1} &\sim&-{iR\over w}\sin\theta\sinh\xi^1=-{2\bar x^0\sin\theta\over w}\\
{{\bar x}^2} &\sim& {iR\over2}
\left(\cos\theta\cosh\xi^1-{2\sin\theta\over w}\right)
={iR\over2}
\left(\cos\theta\sqrt{1-4x^{02}/R^2}-{2\sin\theta\over w}\right)\\ 
\eea
and we see that ${\bar x}_1,{\bar x}_2$ go to $i\infty$ for $w\to0$.
The fact that ${\bar x}_1,{\bar x}_2$ follow a time dependent
(imaginary) trajectory shows that the source is not simply a D-brane.
\section{Worldsheets for AdS and flat space on the lightcone.}
Working in lightcone gauge enables an interesting comparison
of the four-corner scattering solution for 4 strings in AdS$_5$ to that
for 4 strings in flat Minkowski space. Actually, 
for the solutions we are considering $x^3=y^3=0$, so for them 
lightcone time is ordinary time and $p^+$ is ordinary energy.

At first glance the worldsheets in $y$-space describing 
string-string scattering in AdS and flat space 
(compare for example the figure on the right of
 Fig.~\ref{adsws4glue3} to Fig.~\ref{flat4pt}), show
little qualitative difference: they are both similar
surfaces spanning  a polygon of light-like line segments.
To gain more insight into the very different physics in the two cases 
it is helpful to express them in light-cone parameters. 
 \bea \tau(\xi)\equiv i{x^+\over R}
={i\over\sqrt2R}(x^0+x^3)=-{1\over 2\sqrt2}\sinh\xi^1\sinh\xi^0.\eea 
Because $\tau$ satisfies the equation of motion
\bea {\partial\over\partial \xi^0}\left({1\over z^2}{\partial\tau\over\partial\xi^0}\right)+
{\partial\over\partial \xi^1}\left({1\over z^2}
{\partial\tau\over\partial\xi^1}\right)\eea 
we can consistently define $\sigma(\xi)$ by
\bea {\partial\sigma\over\partial\xi^1}
=
{R^2\over z^2}{\partial\tau\over\partial\xi^0},\quad
{\partial\sigma\over\partial\xi^0}
=-
{R^2\over z^2}{\partial\tau\over\partial\xi^1}\eea 
from which we find, after a little manipulation and a particular 
choice of integration constant
\bea \sigma(\xi)={1\over\sqrt2}
\left(\cos\theta\cosh\xi^1-\sin\theta\cosh\xi^0\over\sin\theta\cosh\xi^1
+\cos\theta\cosh\xi^0\right)={1\over \sqrt2 }{y^0\over R}={y^+\over R}.
\eea 
This choice for $\sigma$ is the one that makes the density 
of $p^+$, $\partial y^+/\partial\sigma$ uniform along the string.  
Thus $\tau$ and $\sigma$ 
are the usual parameters of GGRT \cite{goddardgrt}. 
\def\sgn{{\rm sgn\ }}

One can invert the relation between $\xi^0$, $\xi^1$ and $\tau$, $\sigma$ 
by solving quadratic equations:
\bea \tanh^2\xi^0&=&{\sinh \beta+\sqrt{\sinh^2\beta+\delta^2}\over \cosh\beta
+\sqrt{\sinh^2\beta+\delta^2}}\\
\tanh^2\xi^1&=&{-\sinh \beta+\sqrt{\sinh^2\beta+\delta^2}\over \cosh\beta
+\sqrt{\sinh^2\beta+\delta^2}}\eea 
where 
\bea \beta=\log\left({1-\sigma\sqrt2\tan\theta
\over\tan\theta+\sigma\sqrt2}\right),
\quad\delta=-2\sqrt2 \tau.\eea  
The square roots must be taken so that
$(\sgn{\xi^0})(\sgn{\xi^1})=-\sgn{\tau}$.  Plugging these expressions into (33) or (35) then gives the solution in lightcone gauge.

\begin{figure}[ht]
\psfrag{'y1'}{$y^1$}
\psfrag{'y2'}{$y^2$}
\psfrag{'-infty'}{$-\infty$}
\psfrag{'+infty'}{$+\infty$}
\begin{center}
\includegraphics[width=5in]{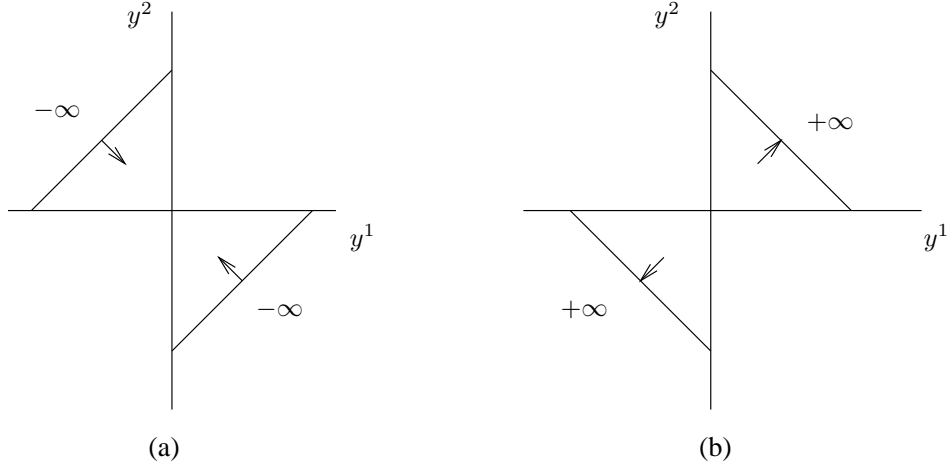}
\caption{Lightcone evolution of two string scattering at 
$90^\circ$. At $\tau=-\infty$
the two strings are parallel straight lines in $y$-space (a). As $\tau$ evolves
the two strings curve with their ends fixed as their centers 
approach the origin. At $\tau=0$ the two strings are stretched along
the coordinate axes with their centers touching at the origin.
At this point they exchange ends and then the centers of the newly
formed strings recede until the strings are again parallel straight
lines at $\tau=+\infty$ (b).}
\label{instrings}
\end{center}
\end{figure}

This physical
parametrization combines information from the worldsheet in $x$ variables
and the T-dual $y$ variables. For example, we can
follow the temporal evolution of the string described by the 
worldsheet solution from very early times $\tau\to-\infty$ to very late
times $\tau\to\infty$.  
For simplicity we take the symmetric case $\theta=\pi/4$, corresponding
to $s=t$.
Then the T-dual coordinates are particularly simple 
\bea
y^1&=&R\sqrt{2}{-\sinh\xi^0\over\cosh\xi_0+\cosh\xi^1},\qquad
y^2=R\sqrt{2}{\sinh\xi^1\over\cosh\xi_0+\cosh\xi^1}.
\eea
We first examine the initial state by taking
$\tau\to-\infty$ at fixed $\sigma$, or $\delta\to\infty$ at fixed $\beta$. 
From the explicit mapping to lightcone parameters we find, as $\delta
\to+\infty$:
\bea
\tanh^2\xi^0&\sim& 1-{1\over\delta}e^{-\beta}+O(\delta^{-2})\\
\tanh^2\xi^1&\sim& 1-{1\over\delta}e^{\beta}+O(\delta^{-2})\\
y^1&\sim& R\sqrt{2}{-\tanh\xi^0\over1+e^\beta}+O(\delta^{-1})\sim
\mp R {1+\sqrt{2}\sigma\over\sqrt{2}}+O(\tau^{-1})\\
y^2&\sim& R\sqrt{2}{\tanh\xi^1\over1+e^{-\beta}}+O(\delta^{-1})\sim
\pm R{1-\sqrt{2}\sigma\over\sqrt{2}}+O(\tau^{-1}),
\eea
where we have taken the signs in the numerator according to
$\sgn\xi^0 \sgn\xi^1=+$ since $\tau<0$. The range of $\sigma$
is $-1<\sigma\sqrt{2}<1$. Then the initial state is two strings
which form parallel straight lines in the $y^1,y^2$ plane. 
For the limit $\tau\to\infty$
the forms are the same except that the relative signs of 
$y^1,y^2$ are reversed. For $\tau=0$ either $y^1=0$ or $y^2=0$.
The evolution of the strings with $\tau$ is indicated
in Fig.~\ref{instrings}.
We see that asymptopia is reached by a power law in $\tau$. 
As we shall soon discover, this is in contrast to the exponential 
approach in flat space. The power law approach is
the signature that the single string in AdS$_5$, 
representing a collection of 
unbound gluons, has a continuous mass spectrum.

Now turning to flat space, we recall that, for the kinematic regime
$s,t<0$ considered here, the lightcone-gauge
flat-space imaginary $x^+\equiv -i\tau$ 
solution is easily obtained by the conformal 
mapping of the lightcone string diagram to the upper half plane
\cite{mandelstam}:
\bea
x^k&=&-i{p_1^k\over\pi}\log|z-1|-i{p_2^k\over\pi}\log|z-u|
-i{p_3^k\over\pi}\log|z|\\
y^k&=&-{p_1^k\over\pi}\tan^{-1}{1-x\over y}
-{p_2^k\over\pi}\tan^{-1}{u-x\over y}
+{p_3^k\over\pi}\tan^{-1}{x\over y}\\
\tau+i\sigma &=& {p^+\over\pi}\left(\log(z-1)-\log(z-u)+\log z\right).
\eea 
Here we work in the frame where $p_1^+=-p_2^+=p_3^+=-p_4^+\equiv p^+>0$.
Calling $\zeta=e^{\pi(\tau+i\sigma)/p^+}$ we find that the inverse
mapping is
\beq
z={1\over2}\left(1+\zeta\pm\sqrt{(\zeta+1-2u)^2+4u(1-u)}\right).
\eeq
The lightcone worldsheet for this kinematic regime
is a two-sheeted strip $-\infty<\tau<+\infty$, $0<\sigma< p^+$
joined along a cut from the branch-point $\zeta=2u-1+2i\sqrt{u(1-u)}$.
Since $|\zeta|=1$ at this point the branch-point is
on the $\tau=0$ axis. The cut may be chosen at fixed $\sigma$
and with $\tau>0$.
In flat space $u$ is a modular parameter which is integrated
over the range $0<u<1$. However to compare with the classical
AdS solution this last integral should also be approximated
by its saddle-point, the value of $u$ which extremizes the
action. 

For purposes of a simple comparison, we might as well
consider the symmetric case $s=t$ where the saddle point
must be $u=1/2$, and for which the branch-point
is at $\tau+i\sigma=p^+/2$. (For the AdS solutions $s=t$ corresponds to
the choice $\theta=\pi/4$.)  In this special case, the Minkowski string
solution is
\bea
x^i_+&=&i{p_1^i\over\pi}\log\left|{\zeta-1+\sqrt{1+\zeta^2}\over2}\right|
+i{p_2^i\over\pi}\log\left|{\zeta+\sqrt{1+\zeta^2}\over2}\right|
+i{p_3^i\over\pi}\log\left|{\zeta+1+\sqrt{1+\zeta^2}\over2}\right|\\
x^i_-&=&i{p_1^i\over\pi}\log\left|{\zeta-1-\sqrt{1+\zeta^2}\over2}\right|
+i{p_2^i\over\pi}\log\left|{\zeta-\sqrt{1+\zeta^2}\over2}\right|
+i{p_3^i\over\pi}\log\left|{\zeta+1-\sqrt{1+\zeta^2}\over2}\right|.
\eea
Here $x^i_\pm$ are the solutions on the two sheets, joined on a cut
running from $\tau+i\sigma=ip^+/2$ to $\tau=+\infty$. The asymptotic
scattering region is at $\tau=\mp\infty$, which corresponds to
$\zeta=0,\infty$ and $z=0,u=1/2,1,\infty$. For example, the
behavior as $\tau\to-\infty$ ($\zeta\to0$) is
\bea
x^i_+
&\sim&i{p_1^i\tau\over p^+}
+i{p_1^i+p_2^i\over\pi}\log{1\over2}+
O(e^{\pi\tau/p^+})\\
x^i_-
&\sim& i{p_2^i+p_3^i\over\pi}\log{1\over2}
+i{p_3^i\tau\over p^+}
+O(e^{\pi\tau/p^+}).
\eea
In contrast to the power behaved approach in AdS, in flat space the approach 
to the asymptotic solution is exponential in $\tau$
because the normal mode frequencies are discrete. This simply reflects
the fact that the string confines in flat space but not in AdS.
\section{String Scattering from external sources in flat target space}
We have interpreted the one-corner solution for a worldsheet in AdS as
describing a gluon scattering off a time-independent external source. The
closest flat-space analogue of such a solution is the scattering of
an open string off a $p$-brane. In this section we study two such
possibilities $p=0,1$ and compare them to the AdS case.
\begin{figure}[ht]
\begin{center}
\includegraphics[width=4in,height=3in]{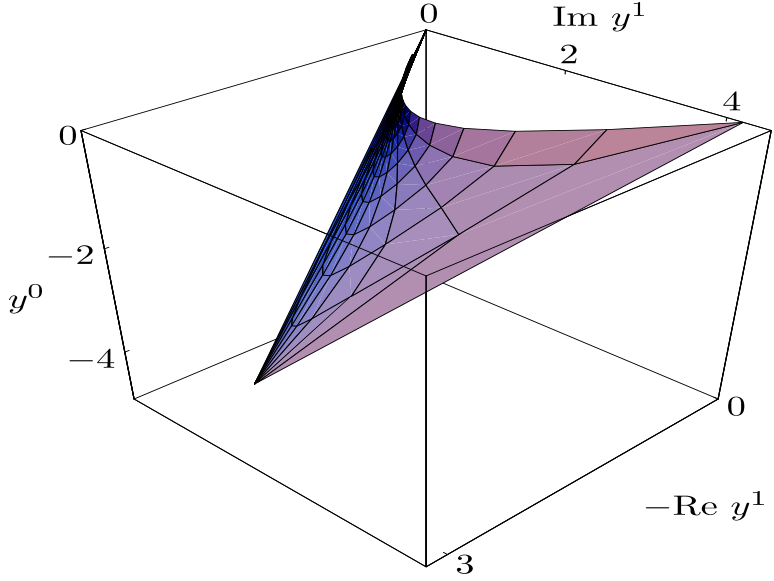}
\caption{The worldsheet in dual $y$ space for the scattering solution of
a string in flat space ending on a 0-brane. The kinematics  is in the
unphysical region where $\cos\theta=(1+u^2)/2u>1$. The case shown is
$u=0.2$ or $\cos\theta=1.04/.4=2.6$. 
The 0-brane sits on the boundary curve in the $y^0=0$ surface. 
The normal derivatives of $y^1,y^2$ to this boundary curve 
in this plane vanish.
The spatial coordinates are complex with $y^2=y^{1*}$.}
\label{0braneu0201}
\end{center}
\end{figure}
\begin{figure}[ht]
\begin{center}
\includegraphics[width=4in,height=3in]{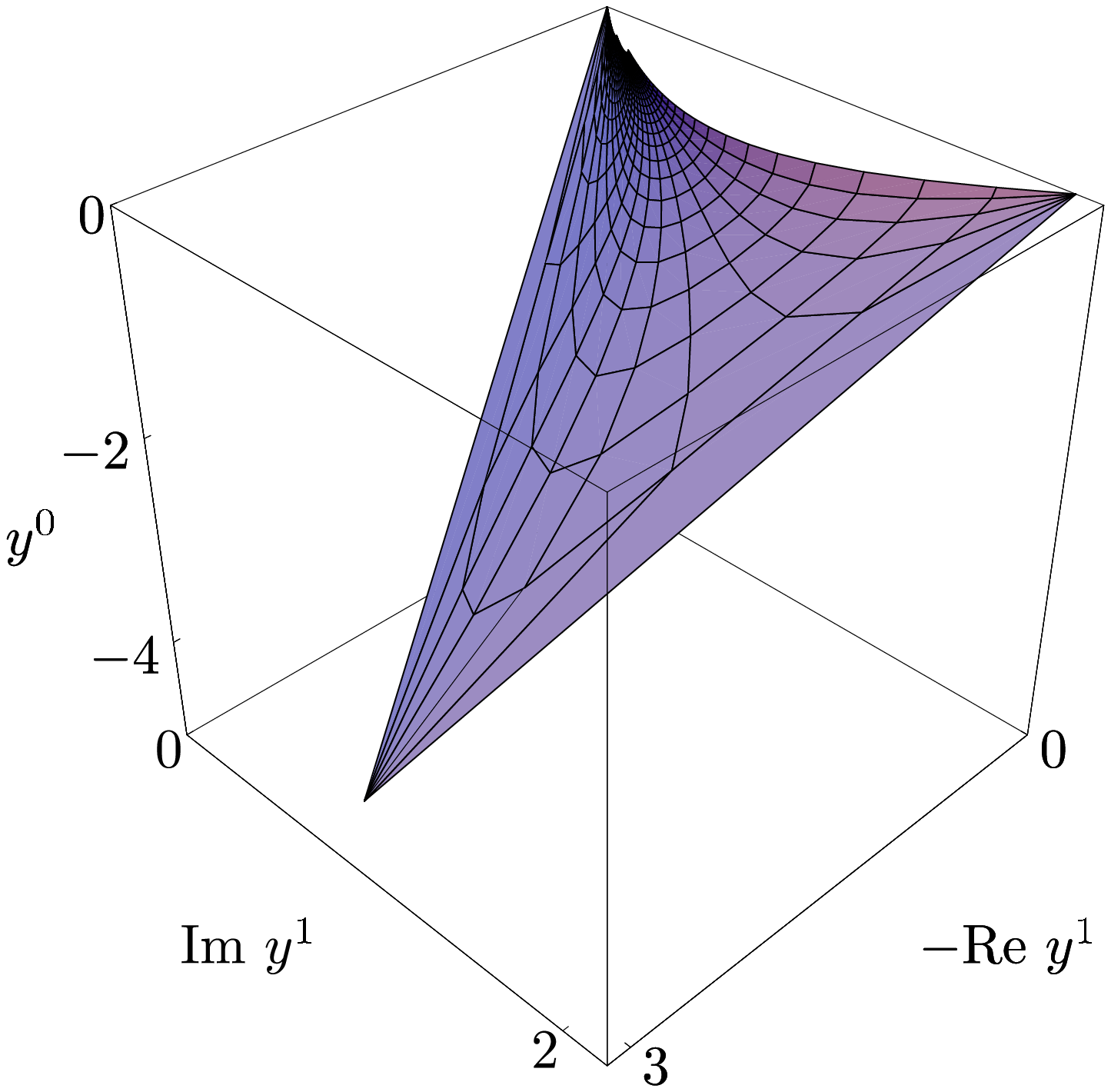}
\caption{The worldsheet in dual $y$ space for the scattering solution of
a string in flat space ending on a 0-brane for $u=0.5$.}
\label{0brane05}
\end{center}
\end{figure}

We first consider a fixed 0-brane source in flat Minkowski space. 
Thus one end of an open string satisfies
Dirichlet conditions on its spatial coordinates
${\vec x}(0,\tau)=0$, but Neumann conditions on its
time component $x^{0\prime}(0,\tau)=0$.
The other end is fully Neumann: $x^{\mu\prime}(\pi,\tau)=0$.
To construct Green functions with mixed boundary conditions it
is convenient to map the worldsheet to the upper right
quadrant of the complex plane. Then for Neumann (Dirichlet) on the $x$-axis
and Dirichlet (Neumann) on the $y$-axis we get
\bea
G_{NN}(z,z^\prime)\!&=&\!\log|z-z^\prime|\!+\!\log|z-z^{\prime*}|
\!+\!\log|z+z^{\prime*}|\!+\!\log|z+z^\prime|\!=\!\log|z^2-z^{\prime2}|\!+\!\log|z^2-z^{\prime*2}|
\\
G_{ND}(z,z^\prime)\!&=&\!\log|z-z^\prime|\!+\!\log|z-z^{\prime*}|
\!-\!\log|z+z^{\prime*}|\!-\!\log|z+z^\prime|\\
G_{DN}(z,z^\prime)\!&=&\!\log|z-z^\prime|\!-\!\log|z-z^{\prime*}|
\!+\!\log|z+z^{\prime*}|\!-\!\log|z+z^\prime|\\
G_{DD}(z,z^\prime)\!&=&\!\log|z-z^\prime|\!-\!\log|z-z^{\prime*}|
\!-\!\log|z+z^{\prime*}|\!+\!\log|z+z^\prime|\!=\!\log|z^2-z^{\prime2}|\!-\!\log|z^2-z^{\prime*2}|.
\eea 
The 0-brane solution, $y^k(z)$ ($x^k(z)$), can be constructed using
$G_{DN}$ ($G_{ND}$) respectively and $y^0(z)$ ($x^0(z)$)
using $G_{DD}$ ($G_{NN}$) respectively. In the first case we impose
for the spatial components,
Dirichlet conditions on the positive real axis and Neumann on the
imaginary axis: $\partial y^k/\partial x=0$.
 $y^k(x,0)=y_1^k$, $0<x<u$; $y^k(x,0)=y_2^k$, $u<x<1$; $y^k(x,0)
=y_3^k$, $1<x<\infty$;
and for the time components Dirichlet conditions on both axes.  This gives
$y^0(x,0)=y_1^0$, $0<x<u$; $y^0(x,0)=y_2^0$, $u<x<1$; $y^0(x,0)=y_3^0$, $1<x<\infty$; $y^0(0,y)=y^0_0$, $0<y<\infty$.
\bea
y^k(x,y)&=&y_3^k+{y_1^k-y_2^k\over\pi}\left[\tan^{-1}{u-x\over y}
+\tan^{-1}{u+x\over y}\right]+{y_2^k-y_3^k\over\pi}\left[\tan^{-1}{1-x\over y}
+\tan^{-1}{1+x\over y}\right]\\
y^0(x,y)&=&y^0_0+2{y^0_1-y^0_0\over\pi}\tan^{-1}{x\over y}
+{y_1^0-y_2^0\over\pi}\left[\tan^{-1}{u-x\over y}
-\tan^{-1}{u+x\over y}\right]\nonumber\\
&&\qquad +{y_2^0-y_3^0\over\pi}\left[\tan^{-1}{1-x\over y}
-\tan^{-1}{1+x\over y}\right]
\eea
We next display the derivatives of the $y$'s:
\bea
{\partial y^k\over\partial x}
&=&{y_1^k-y_2^k\over\pi}\left[{-4uxy\over [y^2+(u-x)^2][ y^2+(u+x)^2]}
\right]+{y_2^k-y_3^k\over\pi}\left[{-4xy\over [y^2+(1-x)^2][y^2+(1+x)^2]}
\right]\\
{\partial y^k\over\partial y}
&=&{y_1^k-y_2^k\over\pi}
\left[{-2u(u^2+y^2-x^2)\over [y^2+(u-x)^2][ y^2+(u+x)^2]}
\right]+{y_2^k-y_3^k\over\pi}
\left[{-2(1+y^2-x^2)\over [y^2+(1-x)^2][y^2+(1+x)^2]}
\right]
\eea
for the spatial components and
\bea
{\partial y^0\over\partial x}
&=&2{y^0_1-y^0_0\over\pi}{y\over y^2+x^2}
+{y_1^0-y_2^0\over\pi}\left[{-2y(y^2+x^2+u^2)\over [y^2+(u-x)^2][y^2+(u+x)^2]}
\right]\nonumber\\
&&\qquad +{y_2^0-y_3^0\over\pi}
\left[{-2y(y^2+x^2+1)\over [y^2+(1-x)^2][y^2+(1+x)^2]}
\right]\\
{\partial y^0\over\partial y}
&=&2{y^0_1-y^0_0\over\pi}{-x\over y^2+x^2}
+{y_1^0-y_2^0\over\pi}\left[{-2x(u^2-y^2-x^2)\over [y^2+(u-x)^2][y^2+(u+x)^2]}
\right]\nonumber\\
&&\qquad +{y_2^0-y_3^0\over\pi}
\left[{-2x(1-y^2-x^2)\over [y^2+(1-x)^2][y^2+(1+x)^2]}
\right]
\eea 
for the time components.  

We next impose the conformal constraints. They should only hold on shell
which for the classical case we take to mean $y^0_0-y^0_1=y^0_3-y^0_0=0$
and $({\vec y}_1-{\vec y}_2)^2-(y^0_1-y^0_2)^2
=({\vec y}_2-{\vec y}_3)^2-(y^0_2-y^0_3)^2=0$. That is the initial and final
0-brane have zero energy and the initial and final open string momenta
are light-like. Then
\beq
{\partial {\vec y}\over \partial x}\cdot{\partial {\vec y}\over \partial y}
-{\partial { y}^0\over \partial x}\cdot{\partial {y^0}\over \partial y}
\!=\!{4\over \pi^2}{xy(1+u^2+2y^2-2x^2)\left[
({\vec y}_1-{\vec y}_2)^2(1+u^2)+2u({\vec y}_1-{\vec y}_2)
\cdot ({\vec y}_2-{\vec y}_3)\right]
\over[y^2+(u-x)^2][y^2+(u+x)^2][y^2+(1-x)^2][y^2+(1+x)^2]}.
\eeq
Setting this to zero requires the modulus $u$ to satisfy
\bea
({\vec y}_1-{\vec y}_2)^2(1+u^2)+2u({\vec y}_1-{\vec y}_2)
\cdot ({\vec y}_2-{\vec y}_3)=0.
\eea
Defining the scattering angle by $({\vec y}_1-{\vec y}_2)
\cdot ({\vec y}_2-{\vec y}_3)=-\cos\theta({\vec y}_1-{\vec y}_2)^2$
we see that the roots are
\beq
u_\pm=\cos\theta\pm\sqrt{\cos^2\theta-1}=e^{\pm i\theta}
\eeq
and are real only for $\cos^2\theta>1$. If $\cos\theta>1$, it
follows that $0<u_-<1<u_+$, so $u_-$ is the root in the assumed range\footnote{
If we want to stay in the physical region, with $\cos^2\theta<1$, $u$
lies on the unit circle and we have to reexamine the boundary
conditions. In this case $\tan^{-1}(u\pm x)/y$ will have discontinuous
$y\to0$ behavior at $x=\pm{\rm Re }u$.}.

To recap, the solution with the on-shell condition imposed is
\bea
y^k(x,y)&=&y_3^k+{y_1^k-y_2^k\over\pi}\left[\tan^{-1}{u_--x\over y}
+\tan^{-1}{u_-+x\over y}\right]\nonumber\\
&&\qquad +{y_2^k-y_3^k\over\pi}\left[\tan^{-1}{1-x\over y}
+\tan^{-1}{1+x\over y}\right]\\
y^0(x,y)&=&y^0_0
+{y_1^0-y_2^0\over\pi}\left[\tan^{-1}{u_--x\over y}
-\tan^{-1}{u_-+x\over y}-\tan^{-1}{1-x\over y}
+\tan^{-1}{1+x\over y}\right].
\eea
Finally we evaluate the action on these solutions, but with $u$
not yet fixed. The integrals converge only if the on-shell
conditions $(y_1-y_2)^2=(y_2-y_3)^2=0$ and $y_3^0=y_1^0=y_0^0$
are imposed\footnote{
For completeness we give the action off-shell, where we introduce two
cutoffs: $x^2+y^2\leq R^2$ and a symmetric $\epsilon$
cutoff surrounding the singular points $x=0,u,1$:
\bea
S&=&{2\over\pi}({\vec y}_1-{\vec y}_2)\cdot({\vec y}_2-
{\vec y}_3)\log{1+u\over1-u}+{2\over\pi}
(y_1^0-y_2^0(y_2^0-y_3^0)\log{1-u^2\over4u}\nonumber\\
&&+{1\over\pi}
\left[(y_1-y_2)^2-2(y_1^0-y_2^0)(y_1^0+y_3^0-2y_0^0)\right]\log u
+{1\over\pi}
\left[(y_1-y_2)^2+(y_2-y_3)^2+2(y_1^0-y_3^0)^2\right]\log2\nonumber\\
&&+{1\over\pi}\left[2(y_1^0-y_0^0)^2-(y_1-y_2)^2-(y_2-y_3)^2\right]
\log\epsilon-{2\over\pi}(y_3^0-y_0^0)^2\log R.
\eea
}. In that case
\bea
S&=&{1\over2}\int_0^\infty dxdy\left[
\left({\partial{\vec y}\over\partial x}\right)^2+
\left({\partial{\vec y}\over\partial y}\right)^2-
\left({\partial{ y^0}\over\partial x}\right)^2-
\left({\partial{y^0}\over\partial y}\right)^2\right]\nonumber\\
&=&{2\over\pi}({\vec y}_1-{\vec y}_2)\cdot({\vec y}_2-
{\vec y}_3)\log{1+u\over1-u}-{2\over\pi}
({\vec y}_1-{\vec y}_2)^2\log{1-u^2\over4u}.
\eea
One can easily check that the condition that $S$ is stationary
under variation in $u$ is the same as that obtained by imposing
the conformal constraints. 

To study the 0-brane solutions with $\cos\theta>1$, we have to allow complex
solutions. Let us define initial and final
momenta ${\vec k}={\vec y}_1-{\vec y}_2$, 
${\vec k}^\prime={\vec y}_2-{\vec y}_3$, and assume scattering is
in the 12 plane. Then consider ${\vec k}=(a,a^*)$, ${\vec k}^\prime
=(a^*,a)$. Then ${\vec k}\cdot{\vec k}^\prime=2|a|^2$ and
${\vec k}^2={\vec k}^{\prime2}=a^2+a^{*2}$, and $\cos\theta
=2|a|^2/(a^2+a^{*2})>1$. Writing $a=\alpha+i\beta$, with $0<\beta<\alpha$ 
we have $u_-=(\alpha-\beta)/(\alpha+\beta)$, which we can solve
for $\beta/\alpha = (1-u_-)/(1+u_-)$. Finally we can set ${\vec y}_3=0$
and $y_0^0=0$. Then
\bea
{\rm Re}\  y^1(x,y)&=&{\rm Re}\ y^2(x,y)
={\alpha\over\pi}\left[\tan^{-1}{u_--x\over y}
+\tan^{-1}{u_-+x\over y}-\tan^{-1}{1-x\over y}
-\tan^{-1}{1+x\over y}\right]\\
{\rm Im}\ y^1(x,y)&=&-{\rm Im}\ y^2(x,y)\nonumber\\
&=&{\alpha\over\pi}{1-u_-\over1+u_-}\left[\tan^{-1}{u_--x\over y}
+\tan^{-1}{u_-+x\over y}+\tan^{-1}{1-x\over y}
+\tan^{-1}{1+x\over y}\right]\\
y^0(x,y)&=&{2\alpha\over\pi}{\sqrt{2u_-}\over1+u_-}
\left[\tan^{-1}{u_--x\over y}
-\tan^{-1}{u_-+x\over y}-\tan^{-1}{1-x\over y}
+\tan^{-1}{1+x\over y}\right].
\eea
We see that $\alpha$ simply sets the overall scale, and
$u_-$ determines the scattering angle according to 
$\cos\theta=(1+u_-^2)/2u_-$. To visualize the surface
we can plot $y^0$ versus say ${\rm Re}\ y^1, {\rm Im}\ y^1$.
as  we do in Figs.~\ref{0braneu0201} and \ref{0brane05}. These
figures should be compared to the one corner AdS worldsheet
on the right of Fig.~\ref{adsws4glue11}. They share a single
corner between light-like segments the other ends of which
are connected by a curve at fixed $y^0$. Note however that
in this 0-brane case the end of the string following this
curve is at a fixed position in transverse coordinate space.

It would be desirable to find an AdS worldsheet describing the 
scattering of a gluon off a heavy quark at a fixed location on the
AdS boundary, $z=0$. The fixed static quark would be described by a
string running from the quark location at $z=0$ on a line at
a fixed point in space to $z=\infty$ ($r=0$). The gluon scattering
would take place near $r=0$ far from the end tied to the quark.
Since the string near that end would possess very large tension,
most of the total energy of the string would reside near
the quark where the string would hardly be perturbed by the scattering.
We haven't been able to find such a solution but it is interesting
to study a flat space analogue which shares this feature; Consider
the scattering of an open string off a  
$1$-brane. In this process $y^0$ and a
spatial coordinate, say $y^3$ satisfy $DD$ boundary conditions and the
remaining spatial coordinates $y^1,y^2$ satisfy $DN$ boundary conditions.
Then $y^3$ has the same form as $y^0$:
\bea
y^3(x,y)&=&y^3_0+2{y^3_1-y^3_0\over\pi}\tan^{-1}{x\over y}
+{y_1^3-y_2^3\over\pi}\left[\tan^{-1}{u-x\over y}
-\tan^{-1}{u+x\over y}\right]\nonumber\\
&&\qquad +{y_2^3-y_3^3\over\pi}\left[\tan^{-1}{1-x\over y}
-\tan^{-1}{1+x\over y}\right].
\eea
The conformal constraint now reads:
\beq
({\vec y}_1-{\vec y}_2)^2-2(y_1-y_2)\cdot(y_0-y_1)+
u^2[({\vec y}_2-{\vec y}_3)^2-2(y_2-y_3)\cdot(y_0-y_1)]
+2u({\vec y}_1-{\vec y}_2)
\cdot ({\vec y}_2-{\vec y}_3)=0
\eeq
where it is understood that the scalar product of the two-vector
$y_0-y_1$ with a four-vector $y_1-y_2$ or $y_2-y_3$ uses only
the $3,0$ components of the four-vector.
Here we take on-shell to mean $(y_1-y_2)^2=(y_2-y_3)^2=0$
(full 4-vector scalar product) and $(y_0-y_1)^2=(y_3-y_0)^2=0$
(2-vector scalar product involving $0,3$ components).
\begin{figure}[ht]
\begin{center}
\includegraphics[width=3in,height=2.25in]{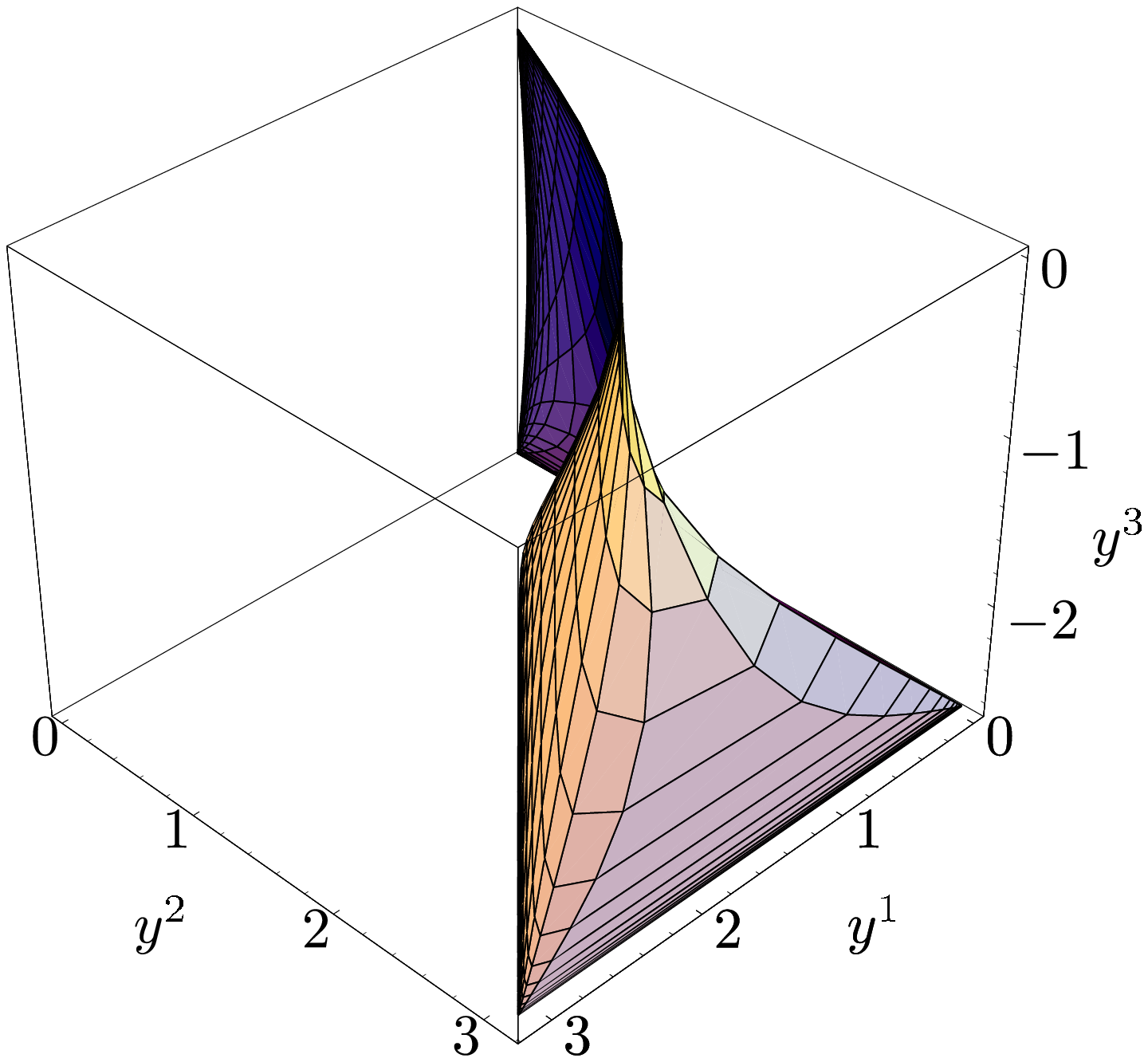}
\quad\includegraphics[width=3in,height=2.25in]{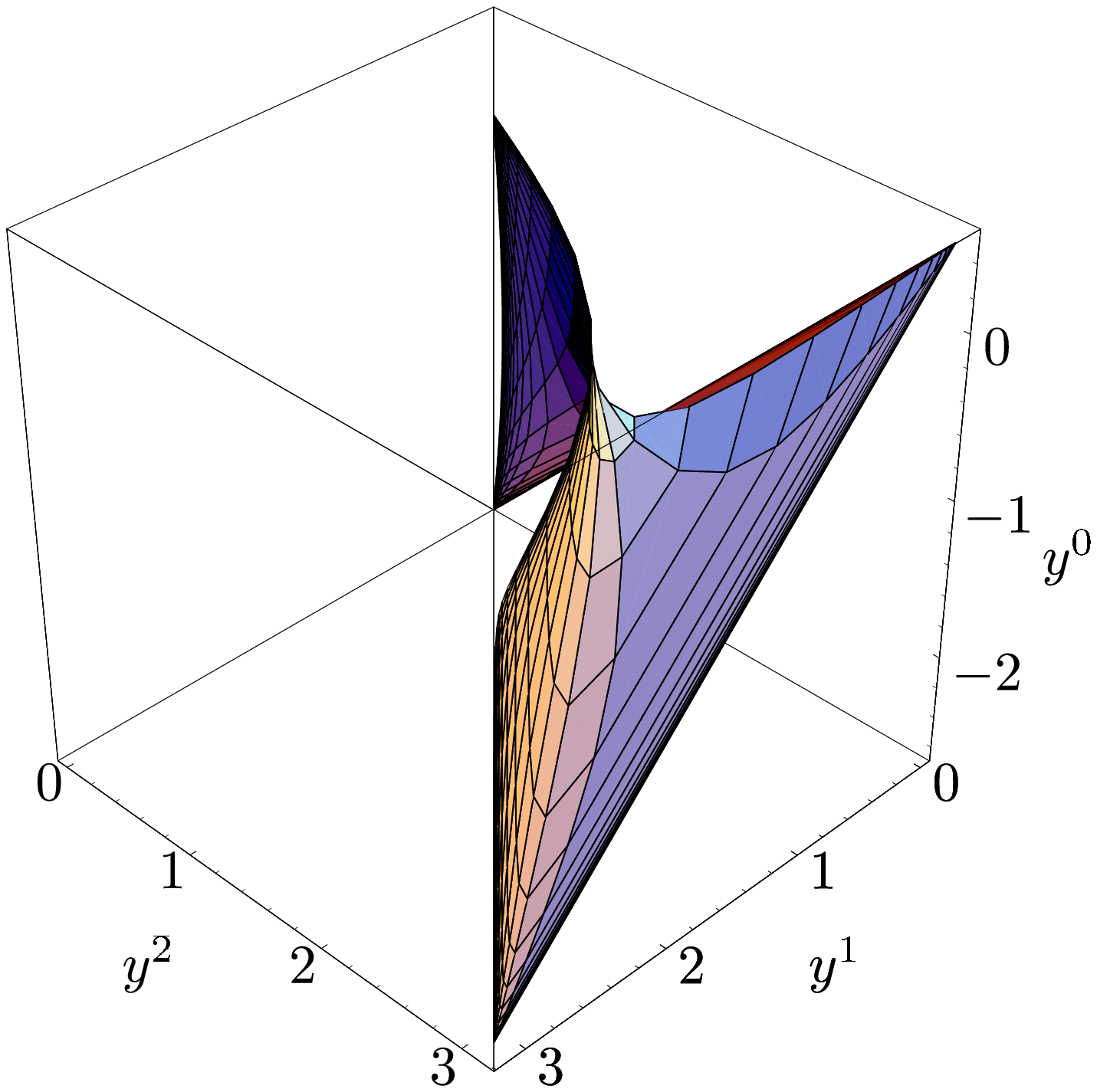}
\caption{The worldsheet in dual $y$ space for the scattering solution of
a string in flat space ending on a 1-brane for $u=0.5$ and $a^\prime=0$. The
left figure shows $y^3$ and the right $y^0$.}
\label{1brane05}
\end{center}
\end{figure}
\begin{figure}[ht]
\begin{center}
\includegraphics[width=3in,height=2.25in]{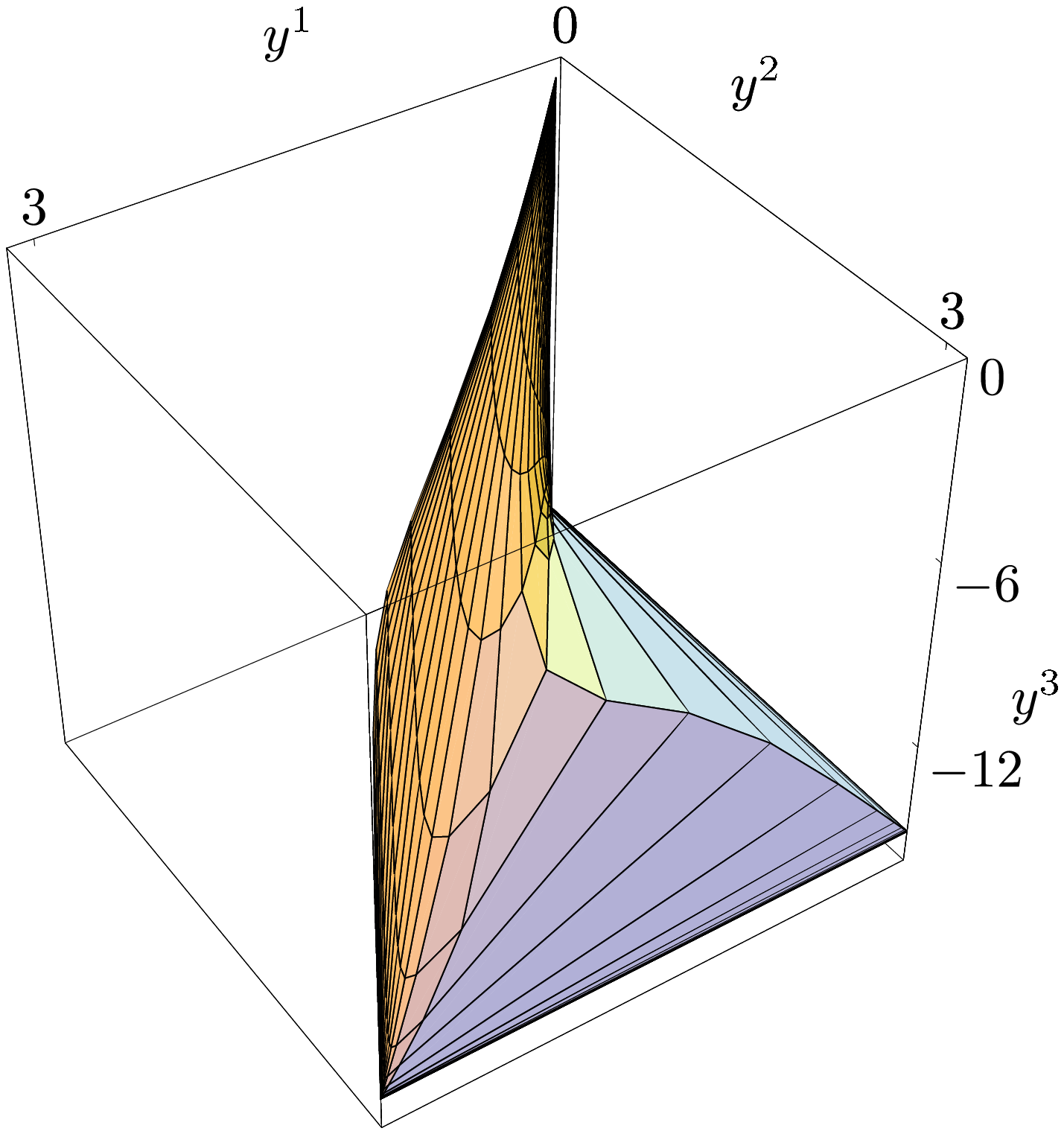}
\quad\includegraphics[width=3in,height=2.25in]{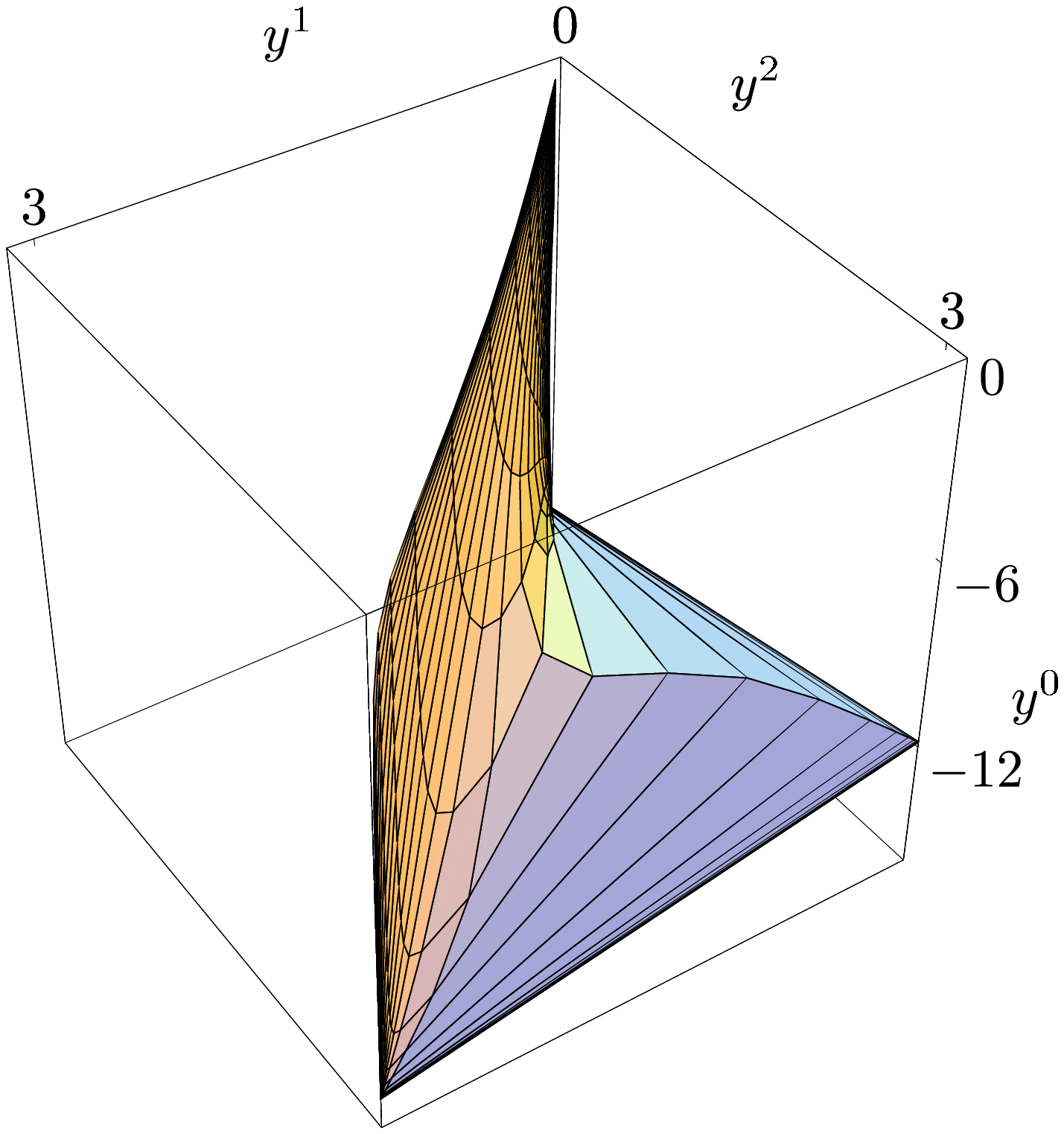}
\caption{The worldsheet in dual $y$ space for the scattering solution of
a string in flat space ending on a 1-brane for $u=0.9$ and $a^\prime=0$.}
\label{1brane09}
\end{center}
\end{figure}

To study the 1-brane solution (see Figs.~\ref{1brane05} and \ref{1brane09}), 
let us take a simplified kinematics,
with $(y_1^3-y_2^3)=(y_2^3-y_3^3)=0$ and $y_0^3-y_1^3=y_0^0-y_1^0
\equiv P=y_3^3-y_0^3=y_3^0-y_0^0$. Let us also take a frame where
$y_1^1-y_2^1=y_2^2-y_3^2=a$ and $y_1^2-y_2^2=y_2^1-y_3^1=a^\prime$.
Then the 1-brane solution reads
\bea
y^1&=&{a\over\pi}\left[\tan^{-1}{u-x\over y}
+\tan^{-1}{u+x\over y}\right]+{a^\prime\over\pi}\left[\tan^{-1}{1-x\over y}
+\tan^{-1}{1+x\over y}\right]\\
y^2&=&{a^\prime\over\pi}\left[\tan^{-1}{u-x\over y}
+\tan^{-1}{u+x\over y}\right]
+{a\over\pi}\left[\tan^{-1}{1-x\over y}
+\tan^{-1}{1+x\over y}\right]\\
y^3&=&-{2P\over\pi}\tan^{-1}{x\over y}\\
y^0&=&-{2P\over\pi}\tan^{-1}{x\over y}-{\sqrt{a^2+a^{\prime2}}\over\pi}
\left[\tan^{-1}{u-x\over y}
-\tan^{-1}{u+x\over y}-\tan^{-1}{1-x\over y}
+\tan^{-1}{1+x\over y}\right].
\eea
The conformal constraint reads
\bea
P={(1+u^2)(a^2+a^{\prime2})+4ua a^\prime\over2\sqrt{a^2+a^{\prime2}}
(1-u^2)}
\eea
and the scattering angle is given by 
$\cos\theta=-2aa^\prime/(a^2+a^{\prime2})$. As a metaphor for the
AdS situation, $y^3$ is simulating the AdS radius $z$ and a 
heavy quark would be simulated by taking $P$ large or $u$ near unity
(see Fig.~\ref{1brane09}).

\vskip14pt

\noindent\underline{Acknowledgments}: 
We would like to thank Martin Kruczenski for very helpful discussions. 
This research was supported in part by the Department
of Energy under Grant No. DE-FG02-97ER-41029.

\appendix
\section{Covariant and lightcone gauges}
In this section we review a variety of possible worldsheet
coordinate choices one can make, 
including conformal and lightcone parameters.
\begin{itemize}
\item Conformal gauge for a Minkowski world sheet
is $g_{00}=-g_{11}$, $g_{01}=0$.
\bea
S_{ws}&\to& \int d\xi^0 d\xi^1 
{R^2T_0\over2z^2}\left[{\dot{{x}}}^2+\dot{z}^2-
{x}^{\prime2}-z^{\prime2}
\right]\\
{\dot x}\cdot x^\prime+{\dot z}\cdot z^\prime&=& {\dot x}^2+x^{\prime2}
+{\dot z}^2+z^{\prime2}=0.
\eea
The equations of motion in this gauge are:
\bea
{\partial\over\partial\xi^0}{{\dot x}^\mu\over z^2}
-{\partial\over\partial\xi^1}{{x}^{\mu\prime}\over z^2} = 0,\qquad
{\partial\over\partial\xi^0}{{\dot z}\over z^2}
-{\partial\over\partial\xi^1}{z^{\prime}\over z^2}+{{\dot x}^2+{\dot z}^2
-x^{\prime2}-z^{\prime2}\over z^3} = 0.
\eea
For a Euclidean worldsheet, $i\xi^0\to\tau$, $\xi_1=\sigma$ and we have
\bea
iS_{ws}&\to& -\int d\tau d\sigma 
{R^2T_0\over2z^2}\left[{\dot{{x}}}^2+\dot{z}^2+
{x}^{\prime2}+z^{\prime2}
\right]\\
{\dot x}\cdot x^\prime+{\dot z}\cdot z^\prime&=& {\dot x}^2-x^{\prime2}
+{\dot z}^2-z^{\prime2}=0\\
{\partial\over\partial\tau}{{\dot x}^\mu\over z^2}
+{\partial\over\partial\sigma}{{x}^{\mu\prime}\over z^2} &=& 0,\qquad
{\partial\over\partial\tau}{{\dot z}\over z^2}
+{\partial\over\partial\sigma}{z^{\prime}\over z^2}+{{\dot x}^2+{\dot z}^2
+x^{\prime2}+z^{\prime2}\over z^3} = 0.
\eea
\item In preparation for lightcone gauge \cite{metsaevtt}, we consider another
family of covariant gauges:
\bea
S_{ws}&\to& \int d\xi^0d\xi^1 
{1\over2}\left[{\dot{{x}}}^2+\dot{z}^2-
{R^4T_0^2\over z^4}({x}^{\prime2}+z^{\prime2})
\right]\\
&&{\dot x}\cdot x^\prime+{\dot z}\cdot z^\prime= {\dot x}^2+{\dot z}^2
+{R^4T_0^2\over z^4}(x^{\prime2} +z^{\prime2})=0.
\eea
The equations of motion in this gauge are:
\bea
{\ddot x}^\mu
-\left({T_0^2R^4\over z^4}{x}^{\mu\prime}\right)^\prime = 0.
\eea
The motivation for this class of gauges
is that the equations of motion are consistent
with identifying $\xi^0$ with a linear combination of the $x^\mu$'s
\item Lightcone time is $x^+=(x^0+x^3)/\sqrt2$.
This leaves ${\bfs x}=(x^1, x^2)$ and $x^-=(x^0-x^3)/\sqrt2$ 
as spatial coordinates. 
Lightcone parametrization of the string means $\xi^0=x^+$ and 
${\cal P}^+=1$, where ${\cal P}^+$ is the momentum conjugate to $x^-$.
Then in this parametrization
\bea
S_{ws}&\to& \int dx^+ \int_0^{p^+} d\sigma 
{1\over2}\left[{\dot{\boldsymbol{x}}}^2+\dot{z}^2-{R^4T_0^2\over z^4}(
{\boldsymbol x}^{\prime2}+z^{\prime2})
\right].\nonumber
\eea
For a closed string one must also impose the constraint
$\int_0^{p^+} d\sigma ({\bfs x}^\prime\cdot{\dot{\bfs{x}}}
+z^\prime \dot{z})=0$.
The equation of motion for ${\bfs x}$ following from this action is 
\bea
{\ddot{\bfs x}}&=&
\left({R^4T_0^2{\bfs x}^\prime\over z^4}\right)^\prime.\nonumber
\eea
\end{itemize}
To put the AdS string action in a form similar to the lightcone
worldsheet action read off from planar graph summation \cite{bardakcit}, 
we do the T-dual transformation
\bea
{\bfs q}^\prime=\dot{\bfs x},\qquad \dot{\bfs q}={R^4T_0^2\over z^4}
{\bfs x}^\prime.\nonumber
\eea
The integrability condition for these equations for ${\bfs q}$ implies
the equation of motion for ${\bfs x}$, and the integrability condition
for the equations for ${\bfs x}$ imply equations of motion for ${\bfs q}$,
which are implied by the worldsheet Lagrangian 
\bea
{\cal L} \to {1\over2}\left[-{\bfs q}^{\prime2}
+{z^4\over R^4T_0^2}\dot{\bfs q}^2
+\dot{z}^2
-{R^4T_0^2\over z^4}z^{\prime2}\right].\nonumber
\eea
Note that the ${\bfs q}$ terms in this Lagrangian have the opposite
sign from simply substituting the dual relations into the $x$ space
Lagrangian. This is necessary to produce the correct equations of
motion, and, as explained in section 2, the correct action does follow
from the proper treatment of duality using the phase space
action principle.
Notice that the $\dot{\bfs q}$ dependence is negligible 
near the boundary of AdS ($z=0$). The intuitive
origin of such terms is explained in the foundational papers
on the lightcone worldsheet \cite{bardakcit,thorngauge,gudmundssontt}. 
We just mention here that a loop is represented 
on the QFT worldsheet by a line segment at
fixed $\sigma$ on which $\dot{\bfs q}=0$. Thus terms in the action
that energetically favor this condition will be gradually brought
into the worldsheet action as one includes more and more loops.
It is very plausible that in the strong 't Hooft coupling limit a
mean field treatment of the sum over loops can be represented
by a bulk term in the action similar to the ${\dot{\bfs q}}^2$
term in the AdS string action.

We can also introduce the dual description in the lightcone friendly
covariant gauge. At the same time it is convenient to use 
$r=R^2/z$ instead of $z$
to represent the AdS radial dimension.
\bea
q^{\mu\prime}={\dot x}^\mu,\qquad 
{\dot q}^\mu={R^4T_0^2\over z^4}x^{\mu\prime}={r^4T_0^2\over R^4}x^{\mu\prime}.
\eea
In these variables the action and reparametrization constraints read
 \bea
S_{ws}&\to& \int d\xi^0d\xi^1 
{1\over2}\left[-{q}^{\prime2}-T_0^2r^{\prime2}+\dot{r}^2{R^4\over r^4}+
{\dot q}^{2}{R^4\over r^4T_0^2}
\right]\\
{\dot q}\cdot q^\prime+T_0^2{\dot r}\cdot r^\prime
&=& {q}^{\prime2}+T_0^2r^{\prime2}+{\dot r}^2{R^4\over r^4}
+{\dot q}^{2}{R^4\over r^4T_0^2}=0.
\eea
We can pass directly to lightcone gauge from here by simply choosing
$q^+=\sigma$, in which case the constraints just turn into
formulas for $q^-$:
\bea
{\dot q}^-={\dot{\bfs q}}\cdot{\bfs q}^\prime+T_0^2{\dot r}\cdot r^\prime,
\qquad
q^{-\prime}={1\over2}\left({\bfs q}^{\prime2}
+T_0^2r^{\prime2}+{\dot r}^2{R^4\over r^4}
+{\dot{\bfs q}}^{2}{R^4\over r^4T_0^2}\right).
\eea
The duality transformation in conformal gauge, for a Euclidean
worldsheet is
\bea
{\dot y}=iR^2{x^\prime\over z^2},\qquad y^\prime=-iR^2{\dot{x}\over z^2},
\eea
where at the same time it is still convenient to use $r=R^2/z$.
Then the Euclidean worldsheet action for the $y$'s is
\bea
S_E=\int d\tau d\sigma {R^2T_0\over2r^2}\left({\dot y}^2+y^{\prime2}+{\dot r}^2
+r^{\prime2}\right).
\eea 
We emphasize that this is {\it not} the Euclidean worldsheet action
for $x,z$ expressed in terms of $y,r$, 
which would have the opposite sign in front of the $y$ terms.
\section{AdS and Conformal Transformations}
\subsection{Poincar\'e Transformations}
Lorentz Transformations $y^{\prime\mu}= \Lambda^{\mu}_{\ \nu}y^\nu$,
$r^\prime=r$:
\bea
Y^{\prime\mu}=\Lambda^{\mu}_{\ \nu}Y^\nu,\qquad Y^{\prime-1}=Y^{-1},
\qquad Y^{\prime4}=Y^4.
\eea   
Translations $y^{\prime\mu}=y^\mu+a^\mu$, $r^\prime=r$:
\bea
Y^{\prime\mu}&=&Y^\mu+a^\mu(Y^{-1}+Y^4)\nonumber\\
Y^{\prime-1}&=&Y^{-1}+a^\mu Y^\mu
+{a^2\over2}(Y^{-1}+Y^4),\qquad Y^{\prime4}=Y^4-a^\mu Y^\mu
-{a^2\over2}(Y^{-1}+Y^4).
\eea
Note that $r^\prime=r$ because $(Y^{-1}+Y^4)$ is invariant under
this transformation.
\subsection{Scaling}
Scaling $y^{\prime\mu}=sy^\mu$, $r^\prime=sr$:
\bea
Y^{\prime\mu}=Y^\mu,\qquad Y^{\prime-1}=\gamma(Y^{-1}+vY^4),\qquad
Y^{\prime4}=\gamma(Y^{4}+vY^{-1})
\eea
with $\gamma=1/\sqrt{1-v^2}$, $s=\sqrt{(1-v)/(1+v)}$.
\subsection{Special Conformal Transformations}
First consider a $04$ boost
\bea
Y^{\prime0}=\gamma(Y^0+vY^4),\qquad Y^{\prime4}=\gamma(Y^4+vY^0)
\eea
translated to Poincar\'e coordinates:
\bea
{1\over r^\prime}&=&{1+\gamma\over2r}+{1-\gamma\over2r}(r^2+y\cdot y)+\gamma v
{y^0\over r}
\nonumber\\
y^{\prime0}&=&{\gamma y^0-\gamma v(r^2+y\cdot y)/2+\gamma v/2\over
(1+\gamma)/2+{(1-\gamma)}(r^2+y\cdot y)/2+\gamma vy^0}
\nonumber\\
y^{\prime k}&=&{y^k\over
(1+\gamma)/2+{(1-\gamma)}(r^2+y\cdot y)/2+\gamma vy^0}.
\eea
We see that the action on the $y$'s is a mixture of translations scaling
and special conformal transformation in the $0$ direction.
We can isolate special conformal transformations by preceding this
boost by a translation and scaling, $y^0=\alpha {\bar y}^0+\beta$,
$y^k=\alpha {\bar y}^k$, $r=\alpha {\bar r}$. $\beta$ is chosen to
remove the constant term in the numerator:
\bea
1+\beta^2+{2\beta\over v}=0,\qquad \beta={\pm\sqrt{1-v^2}-1\over v}.
\eea
Choosing the upper sign leads to
\bea
y^{\prime0}&=&{\alpha {\bar y}^0-\alpha^2\gamma v({\bar r}^2+{\bar y}
\cdot {\bar y})/2\over2(\gamma-1)\sqrt{1-v^2}/\gamma v^2
+\alpha^2{(1-\gamma)}({\bar r}^2+{\bar y}\cdot {\bar y})/2+2(\gamma-1)y^0/v\gamma}
\nonumber\\
y^{\prime k}&=&{\alpha {\bar y}^k\over 2(\gamma-1)\sqrt{1-v^2}/\gamma v^2
+\alpha^2{(1-\gamma)}({\bar r}^2+{\bar y}\cdot {\bar y})/2+2(\gamma-1)y^0/v\gamma}.
\eea
Then choosing $\alpha=2(\gamma-1)\sqrt{1-v^2}/\gamma v^2$ gives
\bea
y^{\prime0}&=&{{\bar y}^0-(\gamma-1)({\bar r}^2+{\bar y}\cdot {\bar y})/v\gamma
\over1
-{(\gamma-1)^2}({\bar r}^2+{\bar y}\cdot {\bar y})/v^2\gamma^2
+2(\gamma-1)y^0/v\gamma}
\nonumber\\
{y}^{\prime k}&=&{{\bar y}^k\over 1
-{(\gamma-1)^2}({\bar r}^2+{\bar y}\cdot {\bar y})/v^2\gamma^2
+2(\gamma-1)y^0/v\gamma}\nonumber\\
{r}^{\prime}&=&{{\bar r}\over 1
-{(\gamma-1)^2}({\bar r}^2+{\bar y}\cdot {\bar y})/v^2\gamma^2
+2(\gamma-1)y^0/v\gamma
}.
\eea
This is seen to be a special conformal transformation
\bea
y^{\prime\mu}={y^\mu-a^\mu y\cdot y\over
1+a\cdot a y\cdot y -2 a\cdot y}
\eea
 on the $y$'s
with $a^\mu=((\gamma-1)/v\gamma, {\vec 0})$ when $r=0$ (Note that
the transformation takes $r=0$ to $r^\prime=0$.)
With $r>0$ it is seen to be a special conformal transformation in the
five dimensional AdS space.

\end{document}